\documentclass[12pt]{article}

\usepackage{graphicx}
\usepackage{latexsym}
\usepackage{epsf}
\usepackage{epsfig}
\usepackage{subfigure}
\usepackage{amssymb,amsmath}
\usepackage[english]{babel}
\usepackage{amsfonts}
\usepackage{amssymb}
\usepackage{graphics}
\usepackage{mathrsfs}
\usepackage{verbatim}
\usepackage{float}
\def\beq{\begin{equation}}
\def\eeq{\end{equation}}
\def\bea{\begin{eqnarray}}
\def\eea{\end{eqnarray}}
\newcommand{\bal}{\begin{align}}
\newcommand{\eal}{\end{align}}
\def\ba{\begin{array}}
\def\ea{\end{array}}
\def\bi{\begin{itemize}}
\def\ei{\end{itemize}}
\def\ben{\begin{enumerate}}
\def\een{\end{enumerate}}
\def\beq{\begin{equation}}
\def\eeq{\end{equation}}
\def\bc{\begin{center}}
\def\ec{\end{center}}
 \def\bt{\begin{table}}
\def\et{\end{table}}
 \def\btb{\begin{tabular}}
\def\etb{\end{tabular}}

\def\mass2{mass${}^2$}

\def\ads{${\mathrm A \mathrm d \mathrm S}_5$}

\def\pa{\partial}

\newcommand{\tr}{\rm Tr}

\newcommand{\cph}{c_\phi}
\newcommand{\sph}{s_\phi}

\newcommand{\ha}{{\hat a}}

\title{
\begin{flushright}
\normalsize{
ANL-HEP-PR-07-101\\
EFI-07-35\\
FERMILAB-PUB-07-630-T\\}
\end{flushright}
 \vspace*{5mm} \Large\textbf{Collider Phenomenology of
Gauge-Higgs Unification Scenarios in Warped Extra Dimensions}
\vspace*{1.0cm}
\author{\textbf{Marcela Carena$^a$, Anibal D.~Medina~$^{b,e}$,
Boris Panes$^{a,f}$,}\\
\textbf{ Nausheen
R.~Shah~$^{c,e}$ and Carlos E.M.~Wagner~$^{c,d,e}$}\\
[0.5cm]
\normalsize\emph{Theoretical Phys. Dept., Fermi National Laboratory,
Batavia, IL 60510, USA~$^a$}\\
\normalsize\emph{Department of Astronomy and Astrophysics~$^b$,
Enrico
Fermi Institute~$^c$}\\
\normalsize\emph{and Kavli Institute
for Cosmological Physics~$^d$,}\\
\normalsize\emph{University of Chicago, 5640 S. Ellis Ave.,
Chicago, IL 60637, USA} \\
\normalsize\emph{HEP Division, Argonne National Laboratory, 9700
Cass Ave., Argonne, IL 60439, USA~$^e$} \\
\normalsize\emph{Dept. de F\'{\i}sica, Univ. Cat\'olica de Chile, Av. V. Mackenna 4860, Santiago, Chile~$^f$}}}
\date{\today}
\begin{document}
\setcounter{page}{0}
\maketitle

\begin{abstract}
We compute the couplings of the zero modes and first excited states of gluons, $W$'s, $Z$ gauge bosons, as well as the Higgs,  to the zero modes
and first excited states of the third generation quarks, in an RS Gauge-Higgs unification scenario based on a bulk $SO(5)\times U(1)_X$ gauge
symmetry, with gauge and fermion fields propagating in the bulk. Using the parameter space consistent with electroweak precision tests and
radiative electroweak symmetry breaking, we study numerically the dependence of these couplings on the parameters of our model. Furthermore,
after emphasizing the presence of light excited states of the top quark, which couple strongly to the Kaluza Klein gauge bosons, the associated
collider phenomenology is analyzed. In particular, we concentrate on the possible detection of the first excited state of the top, $t^1$, which
tends to have a higher mass  than the ones accessible via regular QCD production processes. We stress that the detection of these particles is
still possible  due to an increase in the pair production of $t^1$ induced by the first excited state of the gluon, $G^1$.
\end{abstract}

\thispagestyle{empty}

\newpage

\setcounter{page}{1}

\section{Introduction}

    Five dimensional (5D) warped extra dimensions provide a very attractive beyond
the standard model physics scenario, since the Standard Model (SM) weak scale-Planck scale hierarchy may be explained in a natural way
~\cite{Randall:1999ee}. The observed light quark and lepton masses, as well as the suppression of flavor violating operators is naturally
satisfied provided the quark and gauge fields propagate in the bulk and the first and second generation quark wave functions are suppressed
towards the so-called infrared brane (IR brane), where the Higgs is localized and where the natural scale of energies is of the order of the weak
scale ~\cite{ArkaniHamed:1999dc},\cite{Gherghetta:2000qt},\cite{Huber:2000ie}.

The propagation of gauge and fermion fields in the bulk leads to the mixing of zero-modes with Kaluza Klein (KK) modes, which induces important
tree-level effects on precision electroweak observables~\cite{Davoudiasl:1999tf},\cite{Chang:1999nh}. This happens specially for gauge bosons and
third generation quarks~\cite{Csaki:2002gy}--\cite{Carena:2004zn}. The latter tend to be localized close to the IR brane in order to generate the
large top-quark mass. The introduction of a custodial $SU(2)_R$ symmetry together with a discrete left-right symmetry leads to reduced
corrections to the $T$ parameter ~\cite{Agashe:2003zs},\cite{Agashe:2006at} and helps protect the bottom-quark coupling to the Z gauge boson
against large tree-level corrections. The top-bottom doublet may then be embedded in a bidoublet of the $SU(2)_L \times SU(2)_R$ group. Still,
important corrections to the precision electroweak observables subsist at the one loop-level, and agreement with data for KK masses at the reach
of the LHC may only be obtained in a certain region of fermion mass bulk parameters of the third generation quarks
~\cite{Carena:2006bn},\cite{Carena:2007ua}.

The above requirements may be satisfied in a natural way by embedding the Standard Model gauge $SU(2)_L\times U(1)_{Y}$ group and the global
custodial $SU(2)_R$ group into an $SO(5)\times U(1)_{X}$ gauge symmetry group~\cite{Agashe:2006at}. The $SO(5)\times U(1)_{X}$ symmetry is broken
by boundary conditions at the IR brane down to $SU(2)_L \times SU(2)_R\times U(1)_{X}$ and to $SU(2)_L \times U(1)_{Y}$ at the ultraviolet brane
(UV brane). The five dimensional components of the gauge bosons associated with the broken gauge symmetries at the IR brane have the proper
quantum numbers of the Higgs doublet, leading to a natural implementation of the Gauge-Higgs unification
mechanism~\cite{Agashe:2006at}--\cite{gauge:Higgs:unification}. In Ref.~\cite{Medina:2007hz} the Coleman-Weinberg potential for the Higgs field
was studied and its dependence on the five dimensional mass parameters was determined. It was shown that the region of parameters consistent with
precision electroweak observables is in good agreement with that required to obtain the breakdown of the electroweak symmetry, with the proper
values of the top-quark, bottom-quark and weak gauge boson masses.

One of the most important properties of these type of models is
the existence of light excited states of the top quark. These
quarks are strongly coupled to the gauge bosons KK modes which are
localized towards the IR brane and are light enough so that the
first KK mode of the gluon, $G^1$, tends to decay into them. This,
in turn,  leads to a reduced decay branching ratio of $G^1$ into
top-quarks. These properties, together with an increase in the
width of $G^1$ make the $G^1$ detection via decay into top quarks
more challenging than in the models which have been previously
analyzed in the
literature~\cite{Agashe:2006hk},\cite{Lillie:2007yh},\cite{Lillie:2007ve}.
Moreover, for positive values of the bulk mass parameter
associated with the multiplet of the left-handed top-bottom
doublet which is preferred by flavor
constraints~\cite{Agashe:2006wa}, the couplings of the left- and
right-handed top quarks to $G^1$ become close to each other,
leading to a reduced left-right top-quark production asymmetry. A
similar effect will be present in the LHC phenomenology of the
weak gauge bosons KK modes~\cite{Djouadi:2007eg}.

In this work, we shall concentrate on the production of the first
excited state of the top quark  $t^1$, at the LHC. We shall first
show how to consistently determine the couplings of the quarks to
gauge bosons in a functional way. These couplings will then be
used to compute the decay widths and production cross section of
the first top quark and gluon KK modes. We shall show that the
presence of $G^1$ leads to an important enhancement of the $t^1$
production cross section for masses beyond the ones that can be
tested via direct QCD
production~\cite{Aguilar:2005}--\cite{Witold:2007}. This is very
important, since agreement with precision electroweak observables
tends to be obtained for $t^1$ masses larger than 1~TeV, for which
detection of $t^1$ via QCD production becomes very difficult. We
shall also show that for large $t^1$-masses the branching ratios
for the decay of $t^1$ into a $W^+$ and a bottom-quark,  a  $Z$
and a top-quark, and a top-quark and Higgs boson, are in an
approximate 2:1:1 relation, as required by the Goldstone
equivalence theorem. Moreover, we will show that the technique of
massive jets~\cite{Holdom:2007},\cite{Witold:2007} becomes very
important for the reconstruction of the $t^1$ modes.

This article is organized as follows. In section 2 we describe our 5-dimensional model. The mass spectrum and other relevant parameters of our
model consistent with low energy data were investigated in Ref.~\cite{Medina:2007hz}. In section 3 and 4 we will analyze how to obtain the
properly normalized wave functions for the gauge fields and the fermions, respectively. In section 5 we derive the various couplings necessary to
study the collider phenomenology for this model and we numerically study the dependence of the calculated couplings on the parameters of our
model, as well as the decay branching ratios. In Section 6 we discuss the collider phenomenology. We reserve section 7 for our conclusions.

\section{5-Dimensional Model}
\label{s.5d}

We are interested in a 5D gauge theory with gauge group
$SO(5)\times U(1)_X$. The geometry of our space-time will be that
of RS1~\cite{Randall:1999ee}, with an orbifolded extra spatial
dimension in the interval $x_5 \in [0,L]$. The metric for such a
geometry is given by

\begin{equation} \label{e.wb}
ds^2 = a^2(x_5)\eta_{\mu\nu} dx^\mu dx^\nu - dx_5^2 \, .
\end{equation}
where $a(x_5) = e^{- k x_5}$. The space spanning the fifth dimension
corresponds to a slice of \ads, with branes attached at the two
boundary points: $x_5 = 0$ (UV brane) and $x_5 = L$ (IR brane).

We place our gauge fields, $A_M = A_M^\alpha T^\alpha$ and $B_M$, in the bulk, where $T^\alpha$ are the hermitian generators of the fundamental
representation of $SO(5)$ and generically $\tr[T^{\alpha}.T^{\beta}]=C(5)\delta^{\alpha,\beta}$. The explicit form of the
generators~\cite{Agashe:2004rs} are given in Appendix A. Our fermions $\psi$ also live in the bulk, and they transform under a representation
$t^\alpha$ of $SO(5)$.


The 5D action is
\begin{equation}
S_{5D}= \int d^4 x \int_0^Ldx_5 \sqrt{g} \left (-\frac{1}{4g_5^2}\tr
\{ F_{MN}F^{MN} \} -\frac{1}{4 g_X^2} G_{MN}G^{MN} + \bar{\psi} (i
\Gamma^N D_N- M)\psi \right ),\label{5Daction}
\end{equation}
where $D_N = \pa_N - i A_N^\alpha t^\alpha-iB_N$ and $g_5$ and $g_X$
are the 5D dimensionful gauge couplings.

The choice $C(5)=1$ is a convenient choice, since it allows us to
identify the eigenvalues of our generators as the weak isospin, with
the four dimensional coupling given by $g^2=g_5^2/L$. Any other
choice for $C(5)$ may be absorbed into a redefinition of the gauge
fields or the gauge coupling leaving the physics unchanged.

To construct a realistic 4D low energy theory, we will break the
5D $SO(5)\times U(1)_{X}$ gauge symmetry down to the subgroup
$SO(4)\times U(1)_Y=SU(2)_L\times SU(2)_R\times U(1)_Y$ on the IR
brane and to $SU(2)_L\times U(1)_Y$ on the UV brane, where
$Y/2=T^{3_{R}}+Q_{X}$ is the hypercharge and $Q_{X}$ is the
$U(1)_{X}$ associated charge which is accommodated to obtain the
correct hypercharge. We divide the generators of $SO(5)$ as
follows: the generators of $SU(2)_{L,R}$ are denoted by
$T^{a_{L,R}}$ and $t^{a_{L,R}}$, while the generators from the
coset $SO(5)/SO(4)$ are denoted by $T^\ha$ and $t^\ha$.

In order to obtain the correct hypercharge and therefore the right
Weinberg angle $\theta_{W}$, we need to rotate the fields $A^{
3_{R}}_M\in SU(2)_{R}$ and $B_{M}\in
U(1)_{X}$~\cite{Sakamura:2006rf},
\begin{eqnarray}
\begin{array}{c}
\begin{array}{ccccccccccccc}
\begin{pmatrix}
A^{\prime 3_{\rm R}}_M\\
A^Y_M \end{pmatrix} &=& \begin{pmatrix}
\cph & -\sph \\
\sph & \cph
\end{pmatrix}.\begin{pmatrix}
A^{3_{\rm R}}_M\\
B_M \end{pmatrix}
\end{array}
\end{array}
\end{eqnarray}
\begin{equation}
\cph \equiv \frac{g_5}{\sqrt{g_5^2+g_X^2}} ~~,~~
  \sph \equiv \frac{g_X}{\sqrt{g_5^2+g_X^2}} ~~.
\end{equation}
The correct Weinberg angle is then given by $\sph^2\simeq
\tan^{2}\theta_{W}\simeq(0.23/0.77)\simeq 0.2987$.
We will enforce $A^{Y}_\mu$ to have even parity, corresponding to
the hypercharge gauge boson in the 4D low energy limit. From now on
we will drop the prime on $A^{\prime 3_{\rm R}}$, and it will be
understood that $a_{\rm R}$ refers to $1_{\rm R}, 2_{\rm R}$ and
$\prime 3_{\rm R}$.

To implement the breaking of $SO(5)$ on the two branes as stated
above, we impose the following boundary conditions on the gauge
fields:

\begin{eqnarray}
\pa_5 A_\mu^{a_{\rm L}, Y} = A_\mu^{a_{\rm R}, \ha} =
A_5^{a_{\rm L}, Y} &=& 0 \, , \qquad x_5 = 0 \,\label{bc1}\\
\pa_5 A_\mu^{a_{\rm L}, a_{\rm R}, Y} = A_\mu^{\ha}= A_5^{a_{\rm L},
a_{\rm R}, Y}&=& 0 \, , \qquad x_5 = L \, .\label{bc2}
\end{eqnarray}

As discussed in Ref.~\cite{Medina:2007hz}, this set of boundary conditions on the gauge fields leads to the identification of $A^{\hat{4}}_5$ as
the Higgs field with a non-vanishing vacuum expectation value~(vev).

 We concentrate on the third generation fermions which are the most important for electroweak symmetry
breaking (EWSB) and electroweak precision test (EWPT) considerations. The SM fermions are embedded in full representations of the bulk gauge
group as discussed in~\cite{Carena:2006bn},\cite{Carena:2007ua},\cite{Medina:2007hz}. We therefore introduce in the quark sector three $SO(5)$
multiplets per generation as follows:
\begin{eqnarray}
\label{multiplets}
\begin{array}{c}
\begin{array}{ccccccc}
\xi_{1L} &\sim& Q_{1L} &=& \begin{pmatrix}
\chi^{u}_{1L}(-,+)_{5/3} & q^{u}_L(+,+)_{2/3} \\
\chi^{d}_{1L}(-,+)_{2/3} & q^{d}_L(+,+)_{-1/3}
\end{pmatrix} &\oplus& u^{\prime }_L(-,+)_{2/3}~, \vspace{3mm}
\\
\xi_{2R} &\sim& Q_{2R} &=& \begin{pmatrix}
\chi^{u}_{2R}(-,+)_{5/3} & q^{\prime {u}}_R(-,+)_{2/3} \\
\chi^{d}_{2R}(-,+)_{2/3} & q^{\prime {d}}_R(-,+)_{-1/3}
\end{pmatrix} &\oplus& u_R(+,+)_{2/3}~,
\end{array}
\end{array}
\vspace{3mm}
\end{eqnarray}
\begin{eqnarray}
\begin{array}{l}
\xi_{3R} \sim
\begin{array}{ccccccccccc}
Q_{3R} =
\begin{pmatrix}
\chi^{u}_{3R}(-,+)_{5/3} & q^{\prime \prime u}_R(-,+)_{2/3} \\
\chi^{d}_{3R}(-,+)_{2/3} & q^{\prime \prime d}_R(-,+)_{-1/3}
\end{pmatrix}  \\
\\
\oplus T_{1R} = \begin{pmatrix}
\psi^{\prime }_R(-,+)_{5/3} \\
U^{\prime }_R(-,+)_{2/3} \\
D^{\prime }_R(-,+)_{-1/3} \end{pmatrix} \oplus T_{2R} =
\begin{pmatrix}
\psi^{\prime\prime }_R(-,+)_{5/3} \\
U^{\prime\prime }_R(-,+)_{2/3} \\
D_R(+,+)_{-1/3} \end{pmatrix}, \end{array}
\end{array}\nonumber
\end{eqnarray}
where we show the decomposition under $SU(2)_L\times SU(2)_R$, and
explicitly write the $U(1)_{EM}$ charges. The $Q$s are
bidoublets of $SU(2)_L\times SU(2)_R$, with $SU(2)_L$ acting
vertically and $SU(2)_R$ acting horizontally. $T_{1}$ and
$T_{2}$ transform as $({\bf 3}, {\bf 1})$ and $({\bf 1}, {\bf
3})$ under $SU(2)_L\times SU(2)_R$, respectively, while $u$ and
$u^{\prime }$ are $SU(2)_L\times SU(2)_R$ singlets.

As was done in~\cite{Medina:2007hz} we introduce mass mixing
boundary terms,
\begin{equation}
\mathcal{L}_m = 2\delta(x_{5}-L) \Big[ \bar{u}^\prime_L M_{B_1}
u_R + \bar{Q}_{1L} M_{B_2} Q_{3R} + \mathrm{h.c.} \Big] ~,
\label{localizedmasses}
\end{equation}
where $M_{B_1}$ and $M_{B_2}$ are dimensionless masses.  Since the Higgs mixes, amongst other terms, doublets with singlets, we see that with the
current parity assignments for the singlet component of the first multiplet, we would in principle not have positive parity for this coupling at
the IR brane. However, as discussed in~\cite{Medina:2007hz}, the switch in the singlet parity to $(+,+)$ is completely equivalent to taking
$M_{B_{1}}^{2}\rightarrow 1/M_{B_{1}}^{2}$.

We also show the parities on the indicated 4D chirality, where $-$ and $+$ stands for odd and even parity conditions and the first and second
entries in the bracket correspond to the parities in the UV and IR branes respectively. Let us stress that while odd parity is equivalent to a
Dirichlet boundary condition, the even parity is a linear combination of Neumann and Dirichlet boundary conditions, that is determined via the
fermion bulk equations of motion.

The boundary conditions for the opposite chirality fermion multiplet can be read off the ones above by a flip in both chirality and boundary
condition, $(-,+)_L\rightarrow (+,-)_R$ for example. In the absence of mixing among multiplets satisfying different boundary conditions, the SM
fermions arise as the zero-modes of the fields obeying $(+,+)$ boundary conditions. The remaining boundary conditions are chosen so that
$SU(2)_{L} \times SU(2)_{R}$ is preserved on the IR brane and so that mass mixing terms, necessary to obtain the SM fermion masses after
electroweak symmetry breaking, can be written on the IR brane. Consistency of the above parity assignments with the original orbifold $Z_2$
symmetry at the IR brane was discussed in Ref.~\cite{Medina:2007hz}.

\section{Gauge Fields}
\label{s.GF}

Solving the equations of motion in the presence of the Higgs vev. $h$ is complicated, as these mix the Neumann and Dirichlet modes. However, 5D
gauge symmetry relates these solutions to solutions with $h = 0$~\cite{Hosotani:2005nz}.  These solutions, which we  generally call
$f_G^\alpha(x_5,0)$, are related to the solutions in the presence of a Higgs vev., $f_G^\alpha(x_5,h)$, via a simple gauge transformations:

\begin{equation}
\label{e.vrt} f_G^\alpha(x_5,h) T^\alpha = \Omega^{-1}(x_5,h)f_G^\alpha(x_5,0) T^\alpha \Omega(x_5,h) ,
\end{equation}
where $\Omega(x_5,h)$ removes the vev $h$:
\begin{equation}
\label{om.tr} \Omega(x_5,h) = \exp\left[-iC_hhT^{4}\int_0^{x_5}dy\,a^{-2}(y)\right] .
\end{equation}
and  $C_{h}$ is the Higgs normalization constant chosen such that the Higgs kinetic term is properly normalized, $C_h = g_5 (\int_0^L
a^{-2})^{-1/2}$. Following the procedure outlined in~\cite{Medina:2007hz} leads to the following wave functions for the gauge fields when $h=0$,
consistent with the UV boundary conditions:

\begin{equation}
\begin{array}{cc}
f_{G,n}^{a_{\rm L}}(x_5,0) = C_{G,n,a_{\rm L}} C(x_5,m_n), \quad & f_{G,n}^\ha(x_5,0) =
C_{G,n,\ha} S(x_5,m_n)\\
&\\
f_{G,n}^{Y}(x_5,0) = C_{G,n,Y} C(x_5,m_n), \quad & f_{G,n}^{a_{\rm R}}(x_5,0) =
C_{G,n,a_{\rm R}}S(x_5,m_n) \\
\end{array}
\end{equation}
where the coefficients $C_{G,n,\alpha}$ are normalization constants and $m_{n}$ is the particular KK mass under consideration. The functions
$C[x_5,m_n]$ and $S[x_5,m_n]$ are the solutions to the equations of motion in the case of a vanishing Higgs vev., with the following initial
conditions: $C(0,z) = 1$, $C'(0,z) = 0$, $S(0,z) = 0, S'(0,z) = z$. They are given by~\cite{Falkowski:2006vi},\cite{Pomarol:1999ad},
\begin{eqnarray} C(x_5,z) &=& \frac{\pi z}{ 2 k} a^{-1}(x_5) \left [
 Y_0 \left ( \frac{z}{k } \right )      J_1 \left ( \frac{z}{ k a(x_5)} \right )
- J_0 \left (\frac{z}{k} \right )     Y_1 \left (\frac{z}{ k
a(x_5)} \right ) \right ]\\
S(x_5,z) &=&  \frac{\pi z}{ 2 k} a^{-1}(x_5) \left [
 J_1 \left ( \frac{z}{ k } \right )      Y_1 \left ( \frac{z}{ k a(x_5)} \right )
- Y_1 \left ( \frac{z}{ k} \right )     J_1 \left (\frac{z}{ k a(x_5)} \right ) \right ]
\end{eqnarray}

 We can now calculate $f_{G,n}^\alpha(x_5,h)$, the wave functions with
$h\neq0$, using Eq.~(\ref{e.vrt}). The explicit expressions for $f_G^\alpha(x_5,h)$ are given in the appendix. The IR boundary conditions give us
a system of algebraic equations for the coefficients $C_{G,n,\alpha}$. This system of equations can be broken into four subsets of dependent
equations according to charge. The directions in internal space $\hat{1}, 1_L$ and $1_R$ and $\hat{2}, 2_L$ and $2_R$ mix and correspond to the
components of $W^{+,-}$:
\begin{eqnarray}
&\sqrt{2}\cos\left[\frac{\lambda_G h }{f_h}\right] S[L]
C_{G,\hat{i}}+\sin\left[\frac{\lambda_G h }{f_h}\right] \left(S[L]
C_{G,i_R}-C[L] C_{G,i_L}\right)=0&
\nonumber\\
&&\label{W+1}\\
&2 \left(C_{G,i_L} C'[L]+C_{G,i_R} S'[L]\right)-\sqrt{2}
\sin\left[\frac{\lambda_G h }{f_h}\right] \left(e^{2 k L} h C_h
\left(C[L] C_{G,i_L}-S[L]
C_{G,i_R}\right)-2 C_{G,\hat{i}} S'[L]\right)&\nonumber \\
&+2 \cos\left[\frac{\lambda_G h }{f_h}\right]\left(e^{2 k L} h S[L]
C_h C_{G,\hat{i}}+C_{G,i_L} C'[L]-C_{G,i_R} S'[L]\right)=0&
\nonumber\\
&&\label{W+2}\\
&2\left(C_{G,i_L} C'[L]+C_{G,i_R} S'[L]\right)+\sqrt{2}
\sin\left[\frac{\lambda_G h }{f_h}\right] \left(e^{2 k L} h C_h
\left(C[L] C_{G,i_L}-S[L] C_{G,i_R}\right)-2 C_{G,\hat{i}}
S'[L]\right)&\nonumber \\
&-2 \cos\left[\frac{\lambda_G h }{f_h}\right] \left(e^{2 k L} h S[L]
C_h C_{G,\hat{i}}+C_{G,i_L} C'[L]-C_{G,i_R} S'[L]\right)=0&
\nonumber\\
&&\label{W+3}
\end{eqnarray}
with $i = 1,2$.

The neutral gauge bosons, $Z$ and $\gamma$ are given by a mixture
of $\hat{3}, 3_L, 3_{R'}$ and $Y$:

\begin{eqnarray}
&\sqrt{2}\cos\left[\frac{\lambda_G h }{f_h}\right] S[L]
C_{G,\hat{3}}+\sin\left[\frac{\lambda_G h }{f_h}\right] \left(S[L]
c_{\phi } C_{G,3_R}+C[L]
\left(s_{\phi } C_{G,Y}-C_{G,3_L}\right)\right)=0&\nonumber\\
&&\label{N1}\\
&2 \left(C_{G,3_L}+s_{\phi } C_{G,Y}\right) C'[L]+2 c_{\phi }
C_{G,3_R} S'[L]\nonumber \\
&-\sqrt{2} \sin\left[\frac{\lambda_G h }{f_h}\right] \left(e^{2 k L}
h C_h \left(-S[L] c_{\phi } C_{G,3_R}+C[L] \left(C_{G,3_L}-s_{\phi }
C_{G,Y}\right)\right)-2
C_{G,\hat{3}} S'[L]\right)&\nonumber \\
&+2 \cos\left[\frac{\lambda_G h }{f_h}\right] \left(e^{2 k L} h S[L] C_h C_{G,\hat{3}}+\left(C_{G,3_L}-s_{\phi } C_{G,Y}\right) C'[L]-c_{\phi } C_{G,3_R} S'[L]\right)=0&\nonumber \\
&&\label{N2}
\end{eqnarray}
\begin{eqnarray}
&2 c_{\phi } \left(C_{G,3_L}-s_{\phi } C_{G,Y}\right) C'[L]+2
\left(1+ s_{\phi }^2\right) C_{G,3_R} S'[L]&\nonumber \\
&-\sqrt{2}c_{\phi }  \sin\left[\frac{\lambda_G h }{f_h}\right]
\left(e^{2 k L} h C_h \left(S[L] c_{\phi } C_{G,3_R}+C[L]
\left(-C_{G,3_L}+s_{\phi }
C_{G,Y}\right)\right)+2 C_{G,\hat{3}} S'[L]\right)&\nonumber \\
&+2 c_{\phi } \cos\left[\frac{\lambda_G h }{f_h}\right] \left(-e^{2
k L} h S[L] C_h C_{G,\hat{3}}+\left(-C_{G,3_L}+s_{\phi }
C_{G,Y}\right)
C'[L]+c_{\phi } C_{G,3_R} S'[L]\right)=0& \nonumber\\
&&\label{N3}\\
&2 \left(s_{\phi } C_{G,3_L}+\left(1+c_{\phi }^2\right)
C_{G,Y}\right) C'[L]-2 c_{\phi } s_{\phi } C_{G,3_R} S'[L]
&\nonumber \\
&-\sqrt{2} s_{\phi }\sin\left[\frac{\lambda_G h }{f_h}\right]
\left(e^{2 k L} h C_h \left(S[L] c_{\phi } C_{G,3_R}+C[L]
\left(-C_{G,3_L}+s_{\phi }
C_{G,Y}\right)\right)+2 C_{G,\hat{3}} S'[L]\right)&\nonumber \\
&+2s_{\phi }  \cos\left[\frac{\lambda_G h }{f_h}\right] \left(-e^{2
k L} h S[L] C_h C_{G,\hat{3}}+\left(-C_{G,3_L}+s_{\phi }
C_{G,Y}\right) C'[L]+c_{\phi } C_{G,3_R}
S'[L]\right)=0&\nonumber\\
&&\label{N4}
\end{eqnarray}

Note that in the above, we have dropped the mass dependence of the
gauge boson wave functions, but it should be understood that $m_n$
is taken to be the appropriate mass for the respective modes.

 The procedure to obtain the normalization coefficients will be to drop
one of the equations in the above subsets and solve for the
normalization coefficients in terms of one of them, which can then
be found by normalizing the related gauge field wave functions to
unity. It is a combination of these wave functions which share the
same charge that turns out to be the true wave function for the
particular gauge field under study. For example, explicitly for
the $W^+$, dropping the first equation, we can solve for
$C_{G,\hat{1}}$ and $C_{G,1_R}$ in terms of $C_{G,1_L}$:

\begin{eqnarray}
C_{G,\hat{1}}&=&C_{G,1_L} \frac{-4 \cos\left[\frac{\lambda_G
h}{f_h}\right] C'[L] S'[L]+\sqrt{2} C_h a^{-2}_Lh
\sin\left[\frac{\lambda_G
 h}{f_h}\right] \left(S[L] C'[L]+C[L]
 S'[L]\right)}{
2 S'[L] \left(C_h a^{-2}_Lh \cos\left[\frac{\lambda_G h}{f_h}\right]
S[L]+\sqrt{2} \sin\left[\frac{\lambda_G h}{f_h}\right]
S'[L]\right)},\\
&&\nonumber\\
 C_{G,1_R}&=&-C_{G,1_L}\frac{ C'[L]}{S'[L]}
\end{eqnarray}

where the ``Higgs decay constant'' is defined as
\begin{equation}
f_h^2  =  \frac{1}{g_5^2 \int_0^L dy a^{-2}(y) }\,\label{fh}
\end{equation}
and $\lambda_{G}^2=1/2$.

By normalizing the wave functions which contribute to $W^+$, we find $C_{G,1_L}$ through,

\begin{equation}\label{norm.C1L}
\int^L_0 \left(|f_G^{\hat{1}}(x_5,h)|^2+|f_G^{1_L}(x_5,h)|^2+
|f_G^{1_R}(x_5,h)|^2\right) dx_{5}=1\,
\end{equation}
or equivalently,
\begin{equation}
\int^L_0
\left(|C_{G,\hat{1}}f_G^{\hat{1}}(x_5,0)|^2+|C_{G,1_L}f_G^{1_L}(x_5,0)|^2+
|C_{G,1_R}f_G^{1_R}(x_5,0)|^2\right) dx_{5}=1.
\end{equation}
It is interesting to note that the explicit Higgs dependence in the
normalization equations is always canceled. This is true for the
gauge bosons as well as the fermions. Therefore the only dependence
of these on the Higgs vev is through the normalization coefficients.

The condition that the entire system of equations,
Eqs.~(\ref{W+1})-(\ref{N4}), has a solution is only nontrivially
realized if the determinant is zero. This gives us the spectrum of
 KK masses and was the main focus of Ref.~\cite{Medina:2007hz}. The
normalization coefficients must then be calculated using the procedure defined above. Once all the normalization constants have been computed, we have
all the information necessary to calculate the wave functions for the gauge bosons with the appropriate masses. However, we would like to point
out a subtlety in the neutral sector where after electroweak symmetry breaking we obtain the $Z$ gauge boson and the photon. In principle,
dropping any of the four equations and solving the other three should give us the same result. However, since the Higgs does not couple to the
photon, the photon does not have any component in the $\hat{3}$ and $3_{R}$ directions. Therefore, dropping either $f_G^{\hat{3}}(L,h)=0$ or
$f_G^{3_{R}}(L,h)=0$, does not lead to the photon solution in the massless limit. To compute the coefficients associated with the photon, we need
to drop the boundary condition for either $f_G^{3_L}(L,h)$ or $f_G^{Y}(L,h)$. In this case the $C_{G,\alpha}$ are given by:
\begin{eqnarray}
C_{G,\hat{3}}&=&C_{G,3_L}\frac{C'[L]\left(c_{\phi}^2
a_{L}^{-2}z\sin\left[\frac{\lambda h }{f_h}\right]-2C[L]
S'[L]\cot\left[\frac{\lambda h
}{2f_h}\right]\right)}{\sqrt{2}S'[L]\left(a_{L}^{-2}z\cos\left[\frac{\lambda
h }{f_h}\right]+C[L]S'[L]\right)}\label{gamma1.C}\\
C_{G,3_R}&=&-C_{G,3_L} \frac{c_{\phi } C'[L]}{S'[L]},\label{gamma2.C}\\
C_{G,Y}&=&C_{G,3_L}\frac{\left(a_{L}^{-2}z\cos\left[\frac{\lambda
h }{f_h}\right]-\left(1+s_{\phi}^2-\csc^2\left[\frac{\lambda h
}{2f_h}\right]\right)S[L]C'[L]+C[L]S'[L]\right)}{s_{\phi}\left(a_{L}^{-2}z\cos\left[\frac{\lambda
h }{f_h}\right]+C[L]S'[L]\right)}\label{gamma3.C}
\end{eqnarray}
where we have used the Wronskian relation,
\begin{equation}
S'(x_{5},z)C(x_{5},z)-S(x_{5},z)C'(x_{5},z)=za(x_{5})^{-2}.
\end{equation}
 The consistency of this procedure can be seen by the fact that in this case
in the massless limit $f_G^{\hat{3}}(x_5,h)$, Eq.(\ref{c.4}), and
$f_G^{3'_{R}}(x_{5},h)$, a combination of Eq.(\ref{c.6}) and
Eq.(\ref{c.7}),  are identically zero as we expect for the photon
field.

\section{Fermionic KK profiles}

With the introduction of mass mixing boundary terms, Eq.~(\ref{localizedmasses}), and through the Higgs vev, the different multiplets are now
related via the equations of motion. The details of the calculations for the fermion wave functions and their boundary conditions were discussed
thoroughly in Ref.~\cite{Medina:2007hz}. The fermion vector functions in the $h=0$ gauge and the IR boundary conditions used to calculate the
coefficients $C_{F,i}$ may be determined in a similar way as for the gauge bosons.

Our three fermion multiplets in the $h=0$ gauge can be arranged in vector functions in the following way:
\begin{eqnarray}
\label{f.exp.bc}
\begin{array}{c}
f_{F,1,L}(x_{5},0)=\left[\begin{array}{c} C_{F,1}S_{M_{1}}\\
C_{F,2}S_{M_{1}}\\ C_{F,3}\dot{S}_{-M_{1}}\\
C_{F,4}\dot{S}_{-M_{1}}\\ C_{F,5}S_{M_{1}}\end{array}\right]\\
\\
\\
f_{F,2,R}(x_{5},0)=\left[\begin{array}{c} C_{F,6}S_{-M_{2}}\\
C_{F,7}S_{-M_{2}}\\ C_{F,8}S_{-M_{2}}\\
C_{F,9}S_{-M_{2}}\\
C_{F,10}\dot{S}_{M_{2}}\end{array}\right]
\end{array} & &
f_{F,3,R}(x_{5},0)=\left[\begin{array}{c} C_{F,11}S_{-M_{3}}\\
C_{F,12}S_{-M_{3}}\\ C_{F,13}S_{-M_{3}}\\ C_{F,14}S_{-M_{3}}\\
C_{F,15}S_{-M_{3}}\\ C_{F,16}S_{-M_{3}}\\
C_{F,17}S_{-M_{3}}\\C_{F,18}S_{-M_{3}}\\ C_{F,19}S_{-M_{3}}\\
C_{F,20}\dot{S}_{M_{3}}\end{array}\right]
\end{eqnarray}
where, as for the gauge bosons, the $C_{F,i}$ are normalization
constants. The opposite chiralities have opposite boundary
conditions and can be read from these ones. The solutions to the fermion equations of motion in the h=0 gauge
are written in terms of~\cite{Falkowski:2006vi}:
\begin{equation}
\label{SM} \tilde{S}_M(x_5,z) = \frac{\pi z}{ 2 k} a^{-c-\frac{1}{
2} }(x_5) \left [
 J_{\frac{1}{2}+c} \left (\frac{z}{ k } \right )      Y_{\frac{1}{ 2}+c} \left ( \frac{z}{ k a(x_5)} \right )
- Y_{\frac{1}{2}+c } \left ( \frac{z}{ k } \right )
J_{\frac{1}{2}+c}  \left ( \frac{z}{ k a(x_5)} \right ) \right ],
\end{equation}
which has initial conditions $\tilde{S}_{M}(0,z) = 0$,
$\partial_5\tilde{S}_{\pm M}(0,z) = z$.
The fermion wave function solutions with $h = 0$, Eq.(\ref{f.exp.bc}),
are related to the functions in Eq.(\ref{SM}) via:

\begin{eqnarray}
S_{\pm M}(x_5,z)&=&\frac{e^{\pm M x_5}}{a^{2}(x_5)}\tilde{S}_{\pm M}(x_5,z),\\
&&\nonumber\\
 \dot{S}_{\pm M}(x_5,z)&=& \mp\frac{e^{\pm
Mx_5}}{z a(x_5)}\partial_5\tilde{S}_{\pm M}(x_5,z).
\end{eqnarray}
The normalization of the fermion wave functions for the free fermion case
contains a non-trivial dependence on the warp factor,
\begin{equation} \label{ferm.norm} \int^L_0 dx_5 \;
a(x_5)^3 \;
f^{i}_{F,m_n,(L,R)}(x_{5},0)f^{i}_{F,m_m,(L,R)}(x_{5},0)=\delta_{m,n}.
\end{equation}
where here $f^{i}_{F,m_n,(L,R)}(x_{5},0)$ is proportional to either $S_{M}(x_5,m_n)$ or $\dot{S}_{M}(x_5,m_n)$ and in this case, $m_{n}$ is the
mass spectrum in the free fermion case. The superscript $i$, which denotes the vector component,  is not being summed over and we denote the
chiral indices as L, R for left-handed and right-handed fields. It is therefore convenient to eliminate the dependence on $a(x_5)$ of the
normalization condition by redefining the functions
\begin{equation}
S_{\pm M} \rightarrow a(x_5)^{-3/2} S_{\pm M}, \;\;\;\;
\dot{S}_{\pm M} \rightarrow a(x_5)^{-3/2}  \dot{S}_{\pm M},
\;\;\;\; {\rm or}  \;\;\; f^{i}_{F,m_n,(L,R)} \rightarrow
a(x_5)^{-3/2} f^{i}_{F,m_n,(L,R)}.
\end{equation}
In the following, we shall apply this redefinition, which allows
for a clearer  interpretation of the localization of the fermion
wave functions in the fifth dimension and maintains the validity
of the functional relations given in Eq.~(\ref{f.exp.bc}).

We can write the final fermion vector functions in the presence of
the vev $h$ using the same gauge transformation as for the
gauge bosons:
\begin{eqnarray}
& & f_{F,1,L}(x_{5},h) = \Omega f_{F,1,L}(x_{5},0)\\
& & f_{F,2,R}(x_{5},h) = \Omega f_{F,2,R}(x_{5},0)\\
& & f_{F,3,R}(x_{5},h) = \Omega f_{F,3,R}(x_{5},0)
\end{eqnarray}
Applying the boundary conditions at $x_{5}=L$, taking into account
the mass mixing terms from Eq.~(\ref{localizedmasses}) we derive the
conditions on $f_F(L,h)$:
\begin{eqnarray}
\label{f.IR.bc}
\begin{array}{ccccc}
f_{F,1,R}^{1,...,4} + M_{B_2} f_{F,3,R}^{1,...,4}=0 &\quad&
f_{F,1,R}^{5}+M_{B1}f_{F,2,R}^{5}=0&\quad& f_{F,2,L}^{1,...,4}=0
\\
&&&&\\
 f_{F,3,L}^{1,...,4}-M_{B_2}f_{F,1,L}^{1,...,4}=0 &\quad& f_{F,2,L}^{5}-M_{B_1}
f_{F,1,L}^{5}=0&\quad& f_{F,3,L}^{5,...,10}=0
\end{array}
\end{eqnarray}
where the superscripts denote the vector components.

Asking that the determinant of this system of equations vanishes
so that we get a non-trivial solution, we obtain the spectrum of
the fermion modes~\cite{Medina:2007hz}. We note here that the same
value $\lambda_{F}=1/\sqrt{2}$ is found in the fermionic case as
for gauge bosons.

To find the normalization coefficients for the fermion wave
functions, we use the same procedure as for the gauge bosons:
solving the equations for the boundary conditions at $x_5=L$ and
then normalizing the fermions coupled via the Higgs and boundary
mass mixing terms. The equations are presented in the Appendix. In
the equations as well as in the solutions, the functions are
evaluated at the IR boundary, $x_{5}=L$. For the charge $5/3$
fermions, the coefficients are:

\begin{eqnarray}
&&C_{F,11} = C_{F,1} \frac{M_{B_2} S_{M_1}}{\dot{S}_{-M_3}}\cos\left[\frac{\lambda_F h}{f_h}\right]\label{53.1}\\
\nonumber\\
&&C_{F,15}=C_{F,18}^*=\frac{i}{\sqrt{2}} C_{F,1}\frac{ M_{B_2} S_{M_1}}{\dot{S}_{-M_3}}\sin\left[\frac{\lambda_F h}{f_h}\right]\label{53.2}\\
\nonumber
\end{eqnarray}

Charge $-1/3$ fermions:

\begin{eqnarray}
&&C_{F,14}= C_{F,4}\cos\left[\frac{\lambda_F h}{f_h}\right] \frac{M_{B_2} \dot{S}_{-M_1}}{\dot{S}_{-M_3}}\label{13.1}\\
\nonumber\\
&&C_{F,17}=-\frac{S_{M_3}}{\dot{S}_{-M_3}}C_{F,20}=\frac{i}{\sqrt{2}} C_{F,4} \sin\left[\frac{\lambda_F h}{f_h}\right] \frac{M_{B_2} \dot{S}_{-M_1}}{ \dot{S}_{-M_3}}\label{13.2}\\
\nonumber
\end{eqnarray}

Charge $2/3$ fermions:

\begin{eqnarray}
&&C_{F,3}= C_{F,2}\frac{M_{B_2}^2 S_{M_1} S_{-M_3}+\dot{S}_{M_1} \dot{S}_{-M_3}}{M_{B_2}^2 S_{-M_3} \dot{S}_{-M_1}+S_{-M_1} \dot{S}_{-M_3}}\label{23.1}\\
\nonumber\\
&&C_{F,5}=\sqrt{2}\text{  }C_{F,2}\cot\left[\frac{\lambda_F h}{f_h}\right] \label{23.2}\\
\nonumber\\
&&C_{F,7}=C_{F,8}=\frac{S_{M_2}}{\sqrt{2}\dot{S}_{-M_2}}\tan\left[\frac{\lambda_F
h}{f_h}\right]C_{F,10}\nonumber\\
&&=-\frac{1}{2} C_{F,2}S_{M_2}\frac{ M_{B_2}^2 S_{-M_3} \left(2
\dot{S}_{-M_1} \dot{S}_{M_1}- \sin\left[\frac{\text{$\lambda_F
$h}}{f_h}\right]^2 \left(\dot{S}_{-M_1} \dot{S}_{M_1}-S_{-M_1}
S_{M_1}\right)\right)+2 S_{-M_1} \dot{S}_{M_1} \dot{S}_{-M_3} }{
M_{B_1} \left(\sin\left[\frac{\text{$\lambda_F  $h}}{f_h}\right]^2
S_{-M_2} S_{M_2}+\cos\left[\frac{\lambda_F h}{f_h}\right]^2
\dot{S}_{-M_2} \dot{S}_{M_2}\right) \left(M_{B_2}^2 S_{-M_3}
\dot{S}_{-M_1}+S_{-M_1} \dot{S}_{-M_3}\right)}\nonumber\\
&&\label{23.3}
\end{eqnarray}
\begin{eqnarray}
C_{F,12}&=&-\frac{\sin\left[\frac{\lambda_F
h}{f_h}\right]^2}{1+\cos\left[\frac{\lambda_F
h}{f_h}\right]^2}C_{F,13}=-\frac{1}{2}
C_{F,2}M_{B_2}\sin\left[\frac{\lambda_F h}{f_h}\right]^2\frac{
\dot{S}_{-M_1} \dot{S}_{M_1}- S_{-M_1} S_{M_1}}{M_{B_2}^2
S_{-M_3} \dot{S}_{-M_1}+S_{-M_1} \dot{S}_{-M_3}}\label{23.5}\\
C_{F,16}&=&C^*_{F,19}=\frac{i}{2} C_{F,2}M_{B_2}
\sin\left[\frac{\lambda_F h}{f_h}\right] \cos\left[\frac{\lambda_F
h}{f_h}\right]\frac{
\dot{S}_{-M_1} \dot{S}_{M_1}-S_{-M_1} S_{M_1}}{ M_{B_2}^2 S_{-M_3} \dot{S}_{-M_1}+S_{-M_1} \dot{S}_{-M_3}}\label{23.7}\\
\nonumber
\end{eqnarray}

Again, as for the gauge bosons, we can use the normalization
conditions to solve for the coefficients $C_{F,i}$,
\begin{equation}\label{norm.CF}
\int^L_0\sum_{i}f_{F,i,m_n,(L,R)}^{Q*}(x_5,h).f_{F,i,m_n,(L,R)}^{Q}(x_5,h)dx_{5}=1
\end{equation}
where the index $i$ takes into account the sum over the three different multiplets and the superscript Q denotes the charge of the fermion state.
Here $m_{n}$ denotes the masses of the zero and KK modes fermions of charge Q. We define the vector $f^Q_F(x_5,h)$ as the vector $f_F(x_5,h)$
with all components which do not participate in the state mixing set to zero. Using the equations of motion for the fermion fields and the
boundary conditions associated with them, one can show that once one imposes the normalization condition for one chirality, the other one is
automatically satisfied. We checked the consistency of our analysis by numerically verifying the above statement.
\section{Couplings}

\begin{figure}[ht]

        \centering
        \includegraphics[scale=0.6]{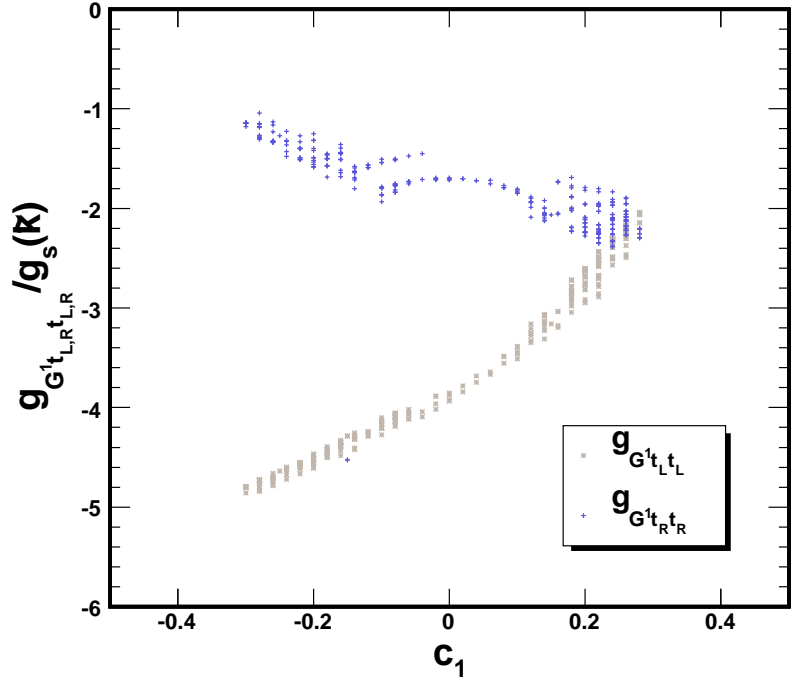}
        \caption{$g_{G^1\bar{t}t}/g_s(\tilde{k})$ vs $c_{1}$. As $c_1$ grows, the coupling for both chiralities unify.}
        \label{collfig1}

\end{figure}
\begin{figure}[ht]

        \centering
        \includegraphics[scale=0.6]{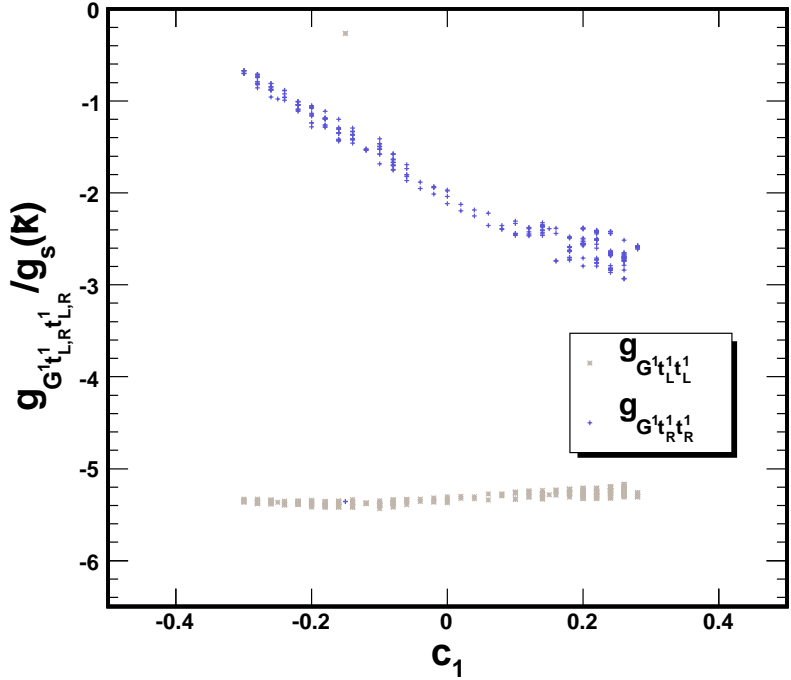}
        \caption{$g_{G^1\bar{t^1}t^1}/g_s(\tilde{k})$ vs $c_{1}$.  As $c_1$ grows, the left-handed coupling remains constant and the right-handed one grows.}
        \label{collfig12}

\end{figure}

\begin{figure}[ht]

        \centering
        \includegraphics[scale=0.6]{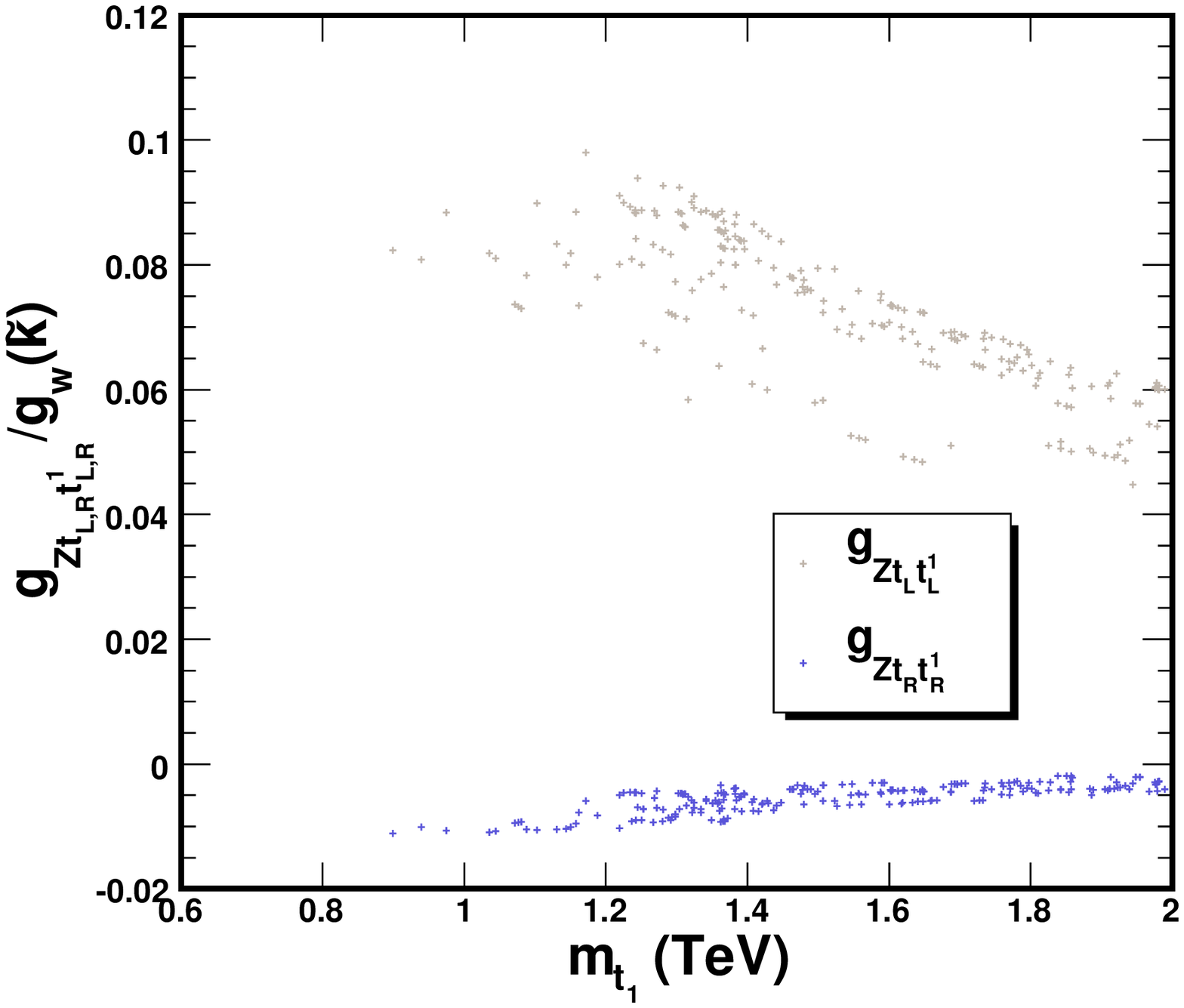}
        \caption{$g_{Z\bar{t^1}t}/g_w(\tilde{k})$ vs $m_{t^1}$ (TeV), where $g_w(\tilde{k})=g_5/\sqrt{L}$. Notice the $1/m_{t^1}$ dependence of $g_{Z\bar{t^1}_Lt_L}$.}
        \label{collfig13}

\end{figure}

Once we have the solutions for all the normalization coefficients, we have the necessary information to compute the various boson-fermion
couplings. We will start with the coupling of the first excited state of the gluon field $G^{1}$ with the fermions. Since gluons do not interact
with the Higgs field, their profile on the fifth dimension is given by $C[x_5,z]$, where in this case $z=m_{G^1}$ is the $G^{1}$-mass.  For
$G^1$-induced pair production of $t^1$'s and for backgrounds to this production mechanism, we are interested in the $g_{G^{1}\bar{t^{1}}t^{1}}$,
$g_{G^{1}\bar{t}t}$ and $g_{G^{1}\bar{t}t^{1}}$ couplings. Since the profiles in the 5th dimension of left-handed and right-handed fields differ,
this leads to left-handed and right-handed couplings which are different. Similar to the description given in Eq.(\ref{5Daction}) for the
electroweak interactions,  these couplings proceed from the covariant derivatives in the fermion kinetic terms,
\begin{eqnarray}\label{Gtt}
g_{G^{1}\bar{t}t}&=&g_{5s}
N_{G^1}\int_0^L\left(\sum_{i}f_{F,i,m_t}^{2/3*}(x_5,h).f_{F,i,m_t}^{2/3}(x_5,h)\right)C[x_5,m_{G^1}]dx_5
\end{eqnarray}
where to simplify notation we omit the chiral subindices.
Furthermore, in the above expression $g_{5s}$ is the 5D strong
coupling related to the 4D strong gauge coupling by
$g_s=g_{5s}/\sqrt{L}$ and $N_{G^1}$ is the normalization for
$G^1$, which is given by
\begin{equation}
N_{G^1}=\left(\int_0^L C^2[x_5,m_{G^1}]dx_{5}\right)^{-1/2}
\end{equation}
 To obtain $g_{G^1\bar{t^{1}}t^{1}}$ and
$g_{G^1\bar{t^{1}}t}$ we only need to replace $m_t$ by the appropriate fermion mass in Eq.(\ref{Gtt}).

Likewise, we calculate the Higgs-fermion coupling
$g_{H\bar{t}t^{1}}$, which will be important when studying the
decay modes of $t^1$ to Higgs and tops. We look again at the
kinetic term for the fermions. The Higgs originates from the
zero-mode of the $A^{\hat{4}}_{5}$ gauge field which enters in the
fermion covariant derivative. Thus the coupling, which mixes left
and right handed fields (left-right or right-left), takes the
form,
\begin{eqnarray}\label{HtT}
g_{H\bar{t}t^{1}}& = &iC_h\int_0^L\left(\sum_{i}f_{F,i,m_{t}}^{*2/3}(x_5,h)f_{H}.T^{\hat{4}}.f_{F,i,m_{t^1}}^{2/3}(x_5,h)\right)e^{-kx_5}dx_5\nonumber\\
& = &i\frac{C_h}{2}C_{F,2,1}C_{F,2,0}^{*}\int_0^L
e^{-kx_{5}}\left(if_{H}f_{F,5,0}^{*}(f_{F,2,1}+f_{F,3,1})-i(f_{F,2,0}^{*}+f_{F,3,0}^{*})f_{H}f_{F,5,1} \right.
\nonumber\\
&+&\left.
if_{H}f_{F,10,0}^{*}(f_{F,7,1}+f_{F,8,1})
-i(f_{F,7,0}^{*}+f_{F,8,0}^{*})f_{H}f_{F,10,1}+\left(f_{F,16,0}^{*}-f_{F,19,0}^{*}\right)\frac{f_{H}f_{F,12,1}}{\sqrt{2}} \right.
\nonumber\\
&+& \left. \left(f_{F,16,0}^{*}-f_{F,19,0}^{*}\right)\frac{f_{H}f_{F,13,1}}{\sqrt{2}}+
\left(f_{F,12,0}^{*}+f_{F,13,0}^{*}\right)\frac{f_{H}f_{F,16,1}}{\sqrt{2}}-\left(f_{F,12,0}^{*}+f_{F,13,0}^{*}\right)\frac{f_{H}f_{F,19,1}}{\sqrt{2}}
\right)\nonumber\\
\end{eqnarray}
where $f_{H}=e^{2kx_{5}}$ is the Higgs profile and $C_{h}$ is the Higgs field normalization. For simplicity we have written $f_{F,i,m_t} =
f_{F,i,0}$,  $f_{F,i,m_{t^1}} = f_{F,i,1}$ and we again omit the chiral indices. The subscripts $0$ and $1$ on the $C's$ and the fermion
functions denote the fermion mass that should be used, $m_{t}$ or $m_{t^1}$ respectively. Using relationships between the fermion normalization
coefficients given in the appendix, this can be written as,

\begin{eqnarray}\label{HtT.2}
g_{H\bar{t}t^{1}}&=&\frac{C_h}{2}\int^L_0dx e^{kx_5}\left(C_{F,3,0}C_{F,5,1}S_{-M_1,0}S_{M_1,1}-\left(C_{F,5,0}C_{F,2,1}-C_{F,2,0}C_{F,5,1}\right)\dot{S}_{M_1,0}S_{M_1,1}\right.\nonumber\\
&+&\left.2\left(C_{F,7,0}C_{F,10,1}S_{-M_2,0}S_{M_2,1}-C_{F,10,0}C_{F,7,1}\dot{S}_{M_2,0}\dot{S}_{-M_2,1}\right)\right.\nonumber\\
&+& \left.i\sqrt{2}\left(C^*_{F,16,0}\left(C_{F,12,1}+C_{F,13,1}\right)+C_{F,5,0}C_{F,3,1}\dot{S}_{M_1,0}\dot{S}_{-M_1,1}+\left(C_{F,12,0}\right.\right.\right.\nonumber\\
&+&\left.\left.\left.C_{F,13,0}\right)C_{F,16,1}\right)S_{-M_3,0}\dot{S}_{-M_3,1}\right)
\end{eqnarray}
 In the above, the only explicit dependence of the couplings
on the Higgs vev  comes through the fermion normalization coefficients.

For the case of $g_{H\bar{t}t}$ or  $g_{H\bar{t^1}t^1}$, the Higgs couplings take a particularly simple form,
\begin{eqnarray}\label{Htt}
g_{H\bar{t}t} &=&\frac{C_h}{2}\int_0^L
e^{kx_5}\left(C_{F,3,0}C_{F,5,0}\left(\dot{S}_{M_1,0}[x_5]\dot{S}_{-M_1,0}[x_5]-S_{M_1,0}[x_5]S_{-M_1,0}[x_5]\right)\right.\nonumber\\
&+&\left.2C_{F,7,0}C_{F,10,0}\left(\dot{S}_{M_2,0}[x_5]\dot{S}_{-M_2,0}[x_5]-S_{M_2,0}[x_5]S_{-M_2,0}[x_5]\right)\right)dx_5 \nonumber\\
&=&-\frac{C_h}{2}\int_0^L e^{2kx_5}\left(C_{F,3,0}C_{F,5,0}+2C_{F,7,0}C_{F,10,0}\right)dx_5
\end{eqnarray}
where we have used the Crowian relation:
\begin{equation}\label{e.cr} -\dot{S}_M(x_5,z)\dot{S}_{-M}(x_5,z) +
S_M(x_5,z) S_{-M}(x_5,z) = 1/a(x_{5}) \,
\end{equation}
In the case of  $g_{H\bar{t^1}t^1}$, we obtain a similar result with the subindex 0 replaced by 1.

Another important coupling which will be of major interest for the reconstruction of the $t^1$ invariant mass, as will be seen in the collider
section, is the $g_{W^{+}\bar{t^1}b}$ coupling which again comes from the covariant kinetic term of the fermion fields, and is given by:
\begin{eqnarray}\label{Wtb}
g_{W^{+}\bar{t^1}b}&=&\sqrt{2}g_5\int_0^L\left(\sum_{i,\alpha}f_{F,i,m_{t^1}}^{2/3*}(x_5,h).\left(f_{G\alpha}(x_5,h)
T^{\alpha}\right).f_{F,i,m_{b}}^{1/3}(x_5,h)\right)dx_5\nonumber\\
&=&g_5C_{G,5}\int_0^L\left[\frac{C_{F,2,1}^{*}C_{F,4,0}}{2}\left(f_{G,8}f_{F,2,1}^{*}f_{F,4,0}-
f_{G,5}f_{F,3,1}^{*}f_{F,4,0}-f_{G,1}f_{F,5,1}^{*}f_{F,4,0}\right.\right.\nonumber\\
&+&\left.\left.if_{G,5}f_{F,12,1}^{*}f_{F,17,0}+f_{G,1}f_{F,13,1}^{*}f_{F,14,0}+\left(-\frac{i}{\sqrt{2}}f_{G,5}f_{F,14,0}+\frac{1}{\sqrt{2}}f_{G,1}f_{F,17,0}\right)f_{F,16,1}^{*}\right.\right.\nonumber\\
&+&\left.\left.i
f_{G,8}f_{F,12,1}^{*}f_{F,14,0}+f_{G,5}f_{F,13,1}^{*}f_{F,20,0}+\left(-\frac{i}{\sqrt{2}}f_{G,5}f_{F,14,0}+\frac{1}{\sqrt{2}}f_{G,8}f_{F,20,0}\right)f_{F,19,1}^{*}\right)\right]\nonumber\\
\end{eqnarray}
where these couplings are left-left and right-right, $g_{5}$ is the $5D$ weak coupling and for simplicity we have written
$f_{G\alpha}(x_5,h)=f_{G,\alpha}$. Here $\alpha$ runs over the generators associated with the $W$ boson: $\hat{1},1_L$ and $1_R$. The subindex 0
refers to the bottom quark and the subindex 1 to $t^1$. Notice that in this case the second multiplet does not contribute to the coupling since
it doesn't mix to form the bottom 5D profile. The first excited state of the bottom fermion has a mass that puts it out of the reach for the LHC.

The $g_{Z\bar{t^1}t}$ coupling, which is also important as a
decay channel for $t^1$, is similarly given by,
\begin{eqnarray}\label{Ztt}
g_{Z\bar{t^1}t}&=&g_5\int_0^L\left(\sum_{i,\alpha}f_{F,i,m_{t^1}}^{*2/3}(x_5,h).\left(f_G^\alpha(x_5,h)
T^{\alpha}\right).f_{F,i,m_{t}}^{2/3}(x_5,h)+h.c\right)dx_5\nonumber\\
&=&g_5C_{G,7}\int_0^L\left[\frac{C_{F,2,0}C_{F,2,1}^{*}}{2}\left(\left(\left(\frac{4g_X}{3g_5}\right)f_{G,11}-f_{G,7}+f_{G,10}\right)f_{F,2,0}f_{F,2,1}^{*}+f_{G,3}f_{F,2,0}f_{F,5,1}^{*}\right.\right.\nonumber\\
&+& \left.\left.
\left(\left(\frac{4g_X}{3g_5}\right)f_{G,11}+f_{G,7}-f_{G,10}\right)f_{F,3,0}f_{F,3,1}^{*}-f_{G,3}f_{F,7,0}f_{F,10,1}^{*}+f_{G,3}f_{F,8,0}f_{F,10,1}^{*}\right.\right.\nonumber\\
&+& \left.\left.
\left(-f_{G,3}f_{F,2,1}^{*}+f_{G,3}f_{F,3,1}^{*}+\left(\frac{4g_X}{3g_5}\right)f_{G,11}f_{F,5,1}^{*}\right)f_{F,5,0}+f_{G,3}f_{F,3,0}f_{F,5,1}^{*}\right.\right.\nonumber\\
&+&\left.\left.\left(\left(\frac{4g_X}{3g_5}\right)f_{G,11}-f_{G,7}+f_{G,10}\right)f_{F,7,0}f_{F,7,1}^{*}+\left(\left(\frac{4g_X}{3g_5}\right)f_{G,11}+f_{G,7}-f_{G,10}\right)f_{F,8,0}f_{F,8,1}^{*}\right.\right.\nonumber\\
&+& \left.\left.
\left(-f_{G,3}f_{F,7,1}^{*}+f_{G,3}f_{F,8,1}^{*}+\left(\frac{4g_X}{3g_5}\right)f_{G,11}f_{F,10,1}^{*}\right)f_{F,10,0}\right.\right.\nonumber\\
&+&\left.\left.\left(\left(\left(\frac{4g_X}{3g_5}\right)f_{G,11}-f_{G,7}+f_{G,10}\right)f_{F,12,1}^{*}+\frac{i}{\sqrt{2}}f_{G,3}f_{F,16,1}^{*}-\frac{i}{\sqrt{2}}f_{G,3}f_{F,19,1}^{*}\right)f_{F,12,0}\right.\right.\nonumber\\
&+& \left.\left.
\left(\left(\left(\frac{4g_X}{3g_5}\right)f_{G,11}+f_{G,7}-f_{G,10}\right)f_{F,13,1}^{*}-\frac{i}{\sqrt{2}}f_{G,3}f_{F,16,1}^{*}+\frac{i}{\sqrt{2}}f_{G,3}f_{F,19,1}^{*}\right)f_{F,13,0}\right.\right.\nonumber\\
&+&\left.\left.\left(\left(\frac{4g_X}{3g_5}\right)f_{G,11}f_{F,16,1}^{*}+\frac{i}{\sqrt{2}}f_{G,3}f_{F,13,1}^{*}-\frac{i}{\sqrt{2}}f_{G,3}f_{F,12,1}^{*}\right)f_{F,16,0}\right.\right.\nonumber\\
&+&\left.\left.\left(\left(\frac{4g_X}{3g_5}\right)f_{G,11}f_{F,19,1}^{*}+\frac{i}{\sqrt{2}}f_{G,3}f_{F,12,1}^{*}-\frac{i}{\sqrt{2}}f_{G,3}f_{F,13,1}^{*}\right)f_{F,19,0}\right)\right]\nonumber\\
\end{eqnarray}
where $\alpha$ runs over $\hat{3},3_L,3_R$, $Y$ and we introduce
$T^Y=Q_{X}I=\frac{2}{3}\frac{g_X}{g_5}\text{I}$.

The functional dependence of the couplings does not change with KK states. The Standard Model couplings may be obtained from the above
expressions by replacing the appropriate  mass. We have evaluated numerically the couplings $g_{Gtt}$, $g_{Htt}$, $g_{Wtb}$ and $g_{Ztt}$ for the
would be zero-mode masses of all the fields and recovered the Standard Model results, as expected, with slight modifications coming from the
mixing with the KK states which becomes smaller as $h$ becomes much smaller than $\tilde{k}$, where $\tilde{k}\equiv ke^{-kL}$ gives the
characteristic scale of the KK masses.

\subsection{Couplings in the linear regime}

The linear regime is defined as the parameter space of our model where $\lambda h/f_h<0.35$. As discussed in Ref.~\cite{Medina:2007hz}, all KK
masses become larger than the SM particle masses in this regime. The fermion solution, Eq.~(\ref{SM}), for $z\ll\tilde{k}$ takes the form:
\begin{equation}\label{lowenergy}
\tilde{S}_{M}\approx z\int^{x_{5}}_{0}
a^{-1}(x_{5})e^{-2My}dy+\mathcal{O}(z^{3})
\end{equation}

Keeping only up to linear order in $h$ and $z$, we can use the last expression in Eq.(\ref{Htt}) for $g_{H\bar{t}t}$ and see that in
this limiting case this coupling reduces to
\begin{equation}
g_{H\bar{t}t}\approx \frac{m_{t}}{h}=\frac{y_{top}}{\sqrt{2}}\,
\end{equation}
recovering the expression for the top Yukawa coupling,
$y_{top}$.
 Similarly, we can obtain analytical expressions for the
 $g_{W^{+}\bar{t^1}_Lb_L}$, $g_{Z\bar{t^1}_Lt_L}$
 and $g_{H\bar{t^1}_{R}t_{L}}$ couplings in the low energy limit
 which will be of importance when analyzing branching ratios for
 the decay of $t^1$. Since the mass of $t^1$ is smaller than
 $\tilde{k}$, one can use an expansion similar to Eq.(\ref{lowenergy}) to obtain the
 previously mentioned couplings.
We find the following approximate expressions
\begin{eqnarray}
g_{W^{+}\bar{t^1}_Lb_L}&\approx &\frac{g^2}{4}
\left(\frac{h}{m_{t^1}}\right) M_{B_{1}}\left(\frac{\tilde{k}}{m_{t^1}}\right)\left(1+\frac{1}{2(1+2c_1)}\left(\frac{m_{t^1}}{\tilde{k}}\right)^2\right)\nonumber\\
&\times&\frac{(1+2c_1)(1+2c_2)\sqrt{kL}\sqrt{(1-2c_1)(3+2c_1)(3+2c_2)}}{\sqrt{\left((1+2c_1)^2(3+2c_1)+M_{B_{1}}^2(1+2c_2)^2(3+2c_2)\right)}}\label{lowzwtb}\\
 g_{Z\bar{t^1}_{L}t_{L}}&\approx &\frac{g_{W^{+}\bar{t^1}_Lb_L}}{\sqrt{2}}\sqrt{1+s^2_{\phi}}\label{lowzzTt}\\
g_{H\bar{t^1}_{R}t_L}&\approx &
\frac{\sqrt{2}m_{t^1}}{gh}g_{W^{+}\bar{t^1}_Lb_L}\label{lowzhTt}
\end{eqnarray}
Notice that the appearance in $g_{W^{+}\bar{t^1}_Lb_L}$ and $g_{Z\bar{t^1}_Lt_L}$ of an additional power of g, as well as the dependence on
$h/m_{t^1}$ has to do with mixing of the charge $2/3$ states through Yukawa coupling interactions proportional to $h$. These approximate
expressions are in good agreement with numerical values obtained for the full couplings.

The decay widths of $G^1$ to fermions and the decay widths of $t^1$ to $Wb$, $Zt$ and $Ht$ are given by,
\begin{eqnarray}
\Gamma_{G^1tt}&=&\frac{m_{G^1}}{48\pi}\sqrt{1-4\frac{m_t^2}{m_{G^1}^2}}\left(g_{G^1\bar{t}_Lt_L}^2+g_{G^1\bar{t}_Rt_R}^2
-(g_{G^1\bar{t}_Lt_L}^2+g_{G^1\bar{t}_Rt_R}^2-6g_{G^1\bar{t}_Lt_L}g_{G^1\bar{t}_Rt_R})\frac{m_t^2}{m_{G^1}^2}\right)\nonumber\\
\\
\Gamma_{Wt^1b}&=&\frac{g_{W^{+}\bar{t^1}_{L}b_{L}}^2}{32\pi}\frac{m_{t^1}^3}{m_W^2}\left(1-3\frac{m_W^4}{m_{t^1}^4}+2\frac{m_W^6}{m_{t^1}^6}\right)\\
\Gamma_{Zt^1t}&=&\frac{1}{8\pi}f(m_{t^1},m_Z,m_{t})\frac{m_{t^1}^2}{m_Z^2}\left(\frac{g_{Z\bar{t^1}_Lt_L}^2+g_{Z\bar{t^1}_Rt_R}^2}{2}\left(1+\frac{m_Z^2}{m_{t^1}^2}+\frac{m_Z^2m_{t}^2}{m_{t^1}^4}+
\frac{m_{t}^4}{m_{t^1}^4}-2\frac{m_{t}^2}{m_{t^1}^2}-2\frac{m_{Z}^4}{m_{t^1}^4}\right)\right.\nonumber\\
&-&\left.6g_{Z\bar{t^1}_Lt_L}g_{Z\bar{t^1}_Rt_R}m_Z^2\frac{m_{t}}{m_{t^1}^3}\right)\\
\Gamma_{Ht^1t}&=&\frac{1}{8\pi
m_{t^1}^2}f(m_{t^1},m_{t},m_H)\left(\frac{(g_{Ht^1_Lt_R}^2+g_{Ht^1_Rt_L}^2)}{2}(m_{t^1}^2+m_{t}^2-m_{H}^2)+2g_{Ht^1_Rt_L}g_{Ht^1_Lt_R}m_{t}m_{t^1}\right)\nonumber\\
\end{eqnarray}
where
\begin{equation}
f(m_1,m_2,m_3)=\frac{1}{2m_{1}}\sqrt{m_{1}^4+m_2^4+m_3^4-2m_1^2m_2^2-2m_1^2m_3^2-2m_2^2m_3^2}
\end{equation}
We are taking the bottom mass to be negligible with respect to the other masses under consideration and we only consider
$g_{W^{+}\bar{t^1}_{L}b_{L}}$ since $g_{W^{+}\bar{t^1}_{R}b_{R}}\ll g_{W^{+}\bar{t^1}_{L}b_{L}}$. For $m_{t^1}\gg m_{t},m_W,m_Z,m_H$, it follows
that $g_{Z\bar{t^1}_{R}t_{R}}\ll g_{Z\bar{t^1}_{L}t_{L}}$, $g_{H\bar{t^1}_{L}t_{R}}\ll g_{H\bar{t^1}_{R}t_{L}}$ and hence the expressions for the
decay widths of $t^1$ reduce to,
\begin{eqnarray}
&& \Gamma_{Wt^1b}=\frac{g_{W^{+}\bar{t^1}_{L}b_{L}}^2}{32\pi}\frac{m_{t^1}^3}{m_W^2}\label{widthw}\\
&& \Gamma_{Z\bar{t^1}t}=\frac{g_{Z\bar{t^1}_Lt_L}^2}{32\pi}\frac{m_{t^1}^3}{m_Z^2}\label{widthZ}\\
&& \Gamma_{Ht^1t}=\frac{g_{Ht^1_Rt_L}^2}{32\pi}m_{t^1}\label{widthH}
\end{eqnarray}
It is easy to see in this limiting case that the Goldstone equivalence theorem applies, by simply replacing in Eqs.(\ref{widthw})--(\ref{widthH})
the analytical expressions for the couplings derived in Eqs.(\ref{lowzwtb})--(\ref{lowzhTt}) . One obtains a relation between the branching
ratios of the decays of $t^1$ into $W^+b$, $Z t$ and $H t$ of approximately 2:1:1, implying that the main decay channel for $t^1$ will be through
$W^+b$.

\begin{figure}[ht]

    \begin{minipage}[b]{0.5\linewidth}

        \centering
        \includegraphics[scale=0.48]{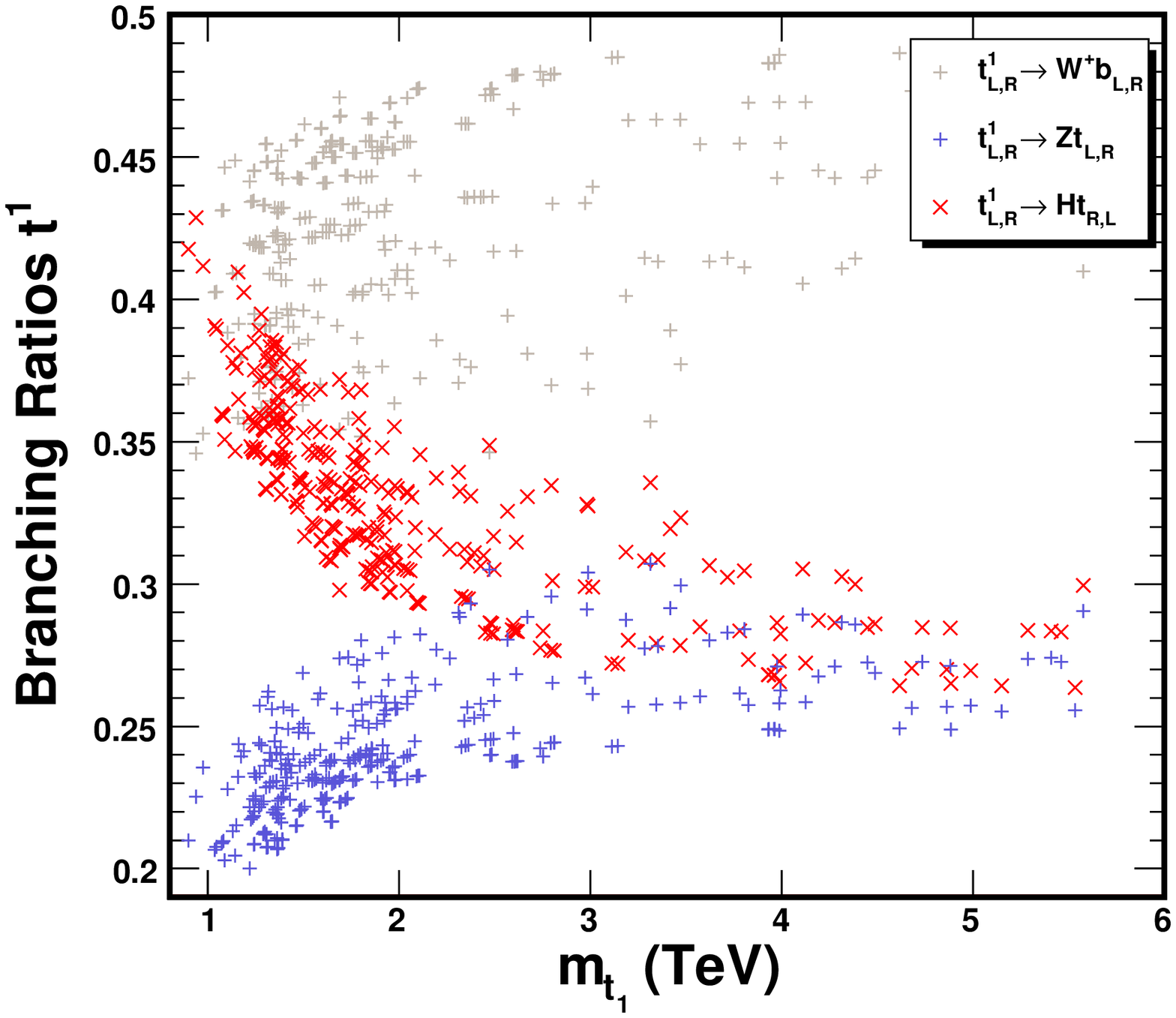}
        \caption{Branching ratios for the decay of $t^1$ vs $m_{t^1}$ (GeV). Notice that the 2:1:1 relations holds for large $m_{t^1}$.}
        \label{collfig14}

  \end{minipage}
    \hspace{0.2cm}
    \begin{minipage}[b]{0.5\linewidth}
\centering
        \includegraphics[scale=0.48]{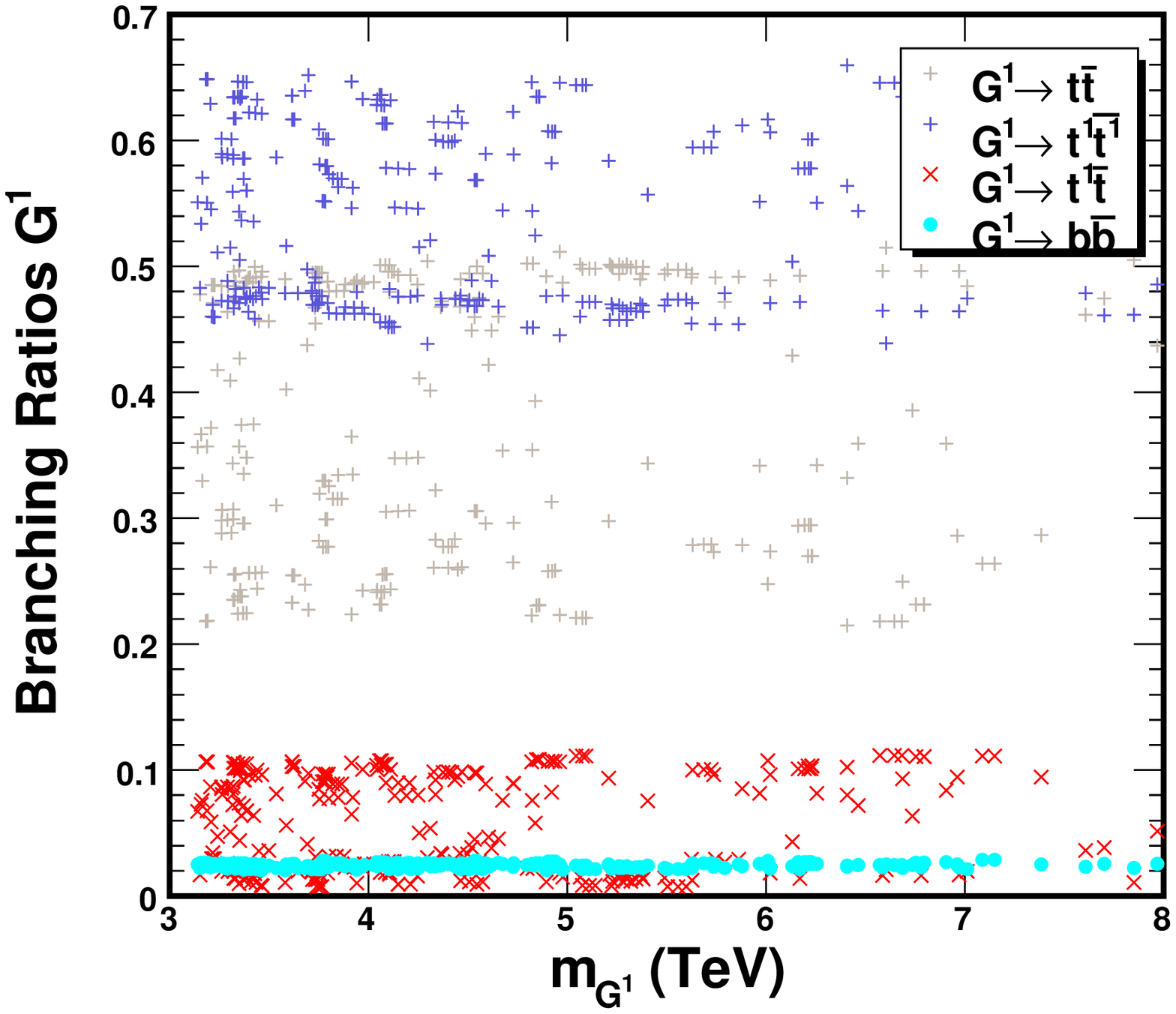}
        \caption{Branching ratios for the decay of $G^1$ vs $m_{G^1}$ (GeV). Notice that $G^1$ decays mostly to $t^1$ pairs.}
        \label{collfig14p}

\end{minipage}

\end{figure}

\subsection{Numerical Results} \label{s.d}
In figures~\ref{collfig1} and~\ref{collfig12} we have plotted the dependence of $g_{G^1\bar{t}t}$ and $g_{G^1\bar{t^1}t^1}$ on the bulk mass
parameter $c_{1}$ which corresponds to the first multiplet. All the couplings were calculated in  the linear regime for values of the parameter
consistent with EWSB and EWPT. In figure~\ref{collfig1} we notice that as $c_{1}$ goes to positive values, the left-handed and right-handed
couplings of the top quark approach each other and take values of the order of two times the strong gauge coupling. Moreover, we can see in
figure~\ref{collfig12} that as $c_{1}$ becomes positive, $g_{G^1\bar{t^1}_Rt^1_R}$ grows becoming similar in size to the approximately common
value of the left and right handed $g_{G^1\bar{t}t}$. Furthermore, $g_{G^1\bar{t^1}_Lt^1_L}$ remains approximately at a constant large value of
about $2g_{G^1\bar{t}t}$ for positive $c_{1}$. In figure~\ref{collfig13} we plot the behavior of the coupling $g_{Z\bar{t^1}t}$ as a function of
$m_{t^1}$. As seen in this figure, the major contribution comes from the left-handed coupling to the fermions. Furthermore, we notice a
dependence of $g_{Z\bar{t^1_{L}}t_{L}}\propto 1/m_{t^1}$, which can be understood by looking at Eq.(\ref{lowzzTt}) and taking into account that
$m_{t^1}\propto \tilde{k}$.
The $g_{W^+\bar{t^1}_{L}b_L}$ coupling shows a similar behavior.

Apart from the couplings, as was previously discussed in Refs.~\cite{Carena:2006bn},\cite{Carena:2007ua},\cite{Medina:2007hz}, a very important
property of this class of models is that the first excited state of the top quark is light enough as to enable the decay of $G^1$ into $t^1$
pairs. This property, together with the large coupling of $G^1$ to $t^1$'s, implies a large branching ratio of the decay of $G^1$ into $t^1$
pairs. In figure~\ref{collfig14} we show the branching ratio for the decay of $t^1$ into $W^+ b$, $Zt$ and $Ht$. In this figure we can appreciate
the appearance of the approximate 2:1:1 relation between the different branching ratios, consistent with the Goldstone equivalence theorem at
large values for the masses of $t^1$. In figure~\ref{collfig14p} we observed, as predicted, the first KK excitation of the gluon decays mostly to
$t^1$ pairs. The subsequent reduction of the branching ratio of the decay of $G^1$ into top-quark pairs,  as well as large width effects, lead to
important modifications of the collider phenomenology with respect to what has been studied in the
literature~\cite{Agashe:2006hk},\cite{Lillie:2007yh},\cite{Lillie:2007ve}, as we will discuss in detail in the next section.

%
%

\section{Collider Phenomenology}

>From the parameter space allowed in this kind of models we see that the typical values for the masses of the first excited gauge bosons and third
generation up-type fermion Kaluza-Klein states are close to 2.5 $\tilde{k}$ and 0.8--1.1 $\tilde{k}$, respectively. Values of  $\tilde{k} \gtrsim
1.4$ TeV lead to consistency with precision electroweak observables and radiative electroweak symmetry breaking in the linear
regime~\cite{Carena:2006bn},\cite{Carena:2007ua},\cite{Medina:2007hz}. This leads to first excited KK gauge bosons with masses larger than about
3.5~TeV, whose wave-functions are localized towards the IR brane. Since first and second generation fermions in these models are localized
towards the UV brane, this implies a reduced coupling of these fermions to $G^1$, which turns out to be on the order of a fifth of the strong
gauge coupling, rendering the search for these heavy $G^1$ states at the LHC  challenging due to the relatively low production cross section.

Most of the previous studies of $G^1$ searches at the LHC have considered the case in which it predominantly decays into top quark
pairs~\cite{Agashe:2006hk},\cite{Lillie:2007yh},\cite{Lillie:2007ve}. As we stressed above, in the class of models analyzed in this article,
which lead to consistency with precision electroweak data and electroweak symmetry breaking, the $G^1$ decays predominantly into $t^1$ pairs.
Moreover, the total decay width of $G^1$ turns out to be in the range of 30 to 40 percent of the $G^1$ mass, which makes it a broad resonance
leading to a harder analysis for the reconstruction of the $G^1$ mass. Therefore, in this work we shall analyze the search for the first excited
fermion state of the top quark and concentrate on the relatively narrow $t^1$ mass reconstruction.

\begin{figure}[ht]

        \centering
        \includegraphics[scale=0.6]{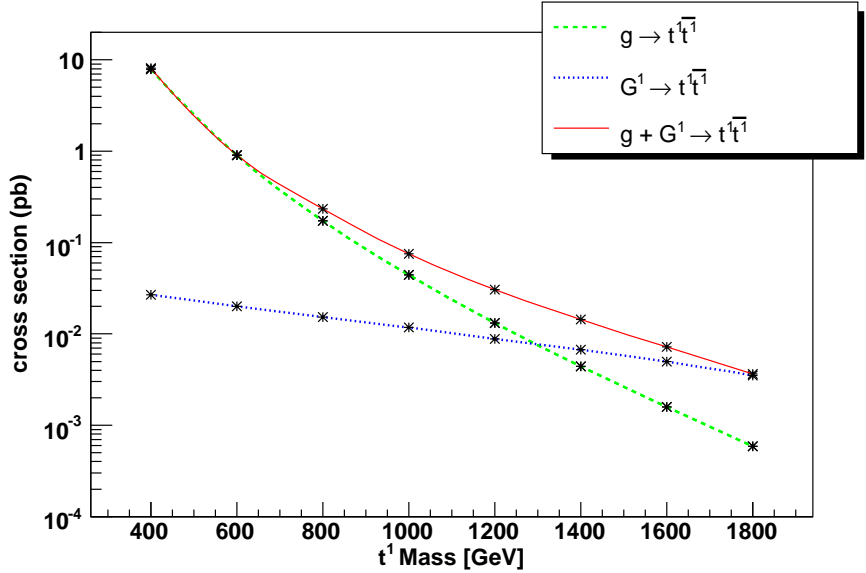}
        \caption{Cross section production at the LHC for  $m_{G^1} = 4.0$ TeV with
                  couplings $g_{G^1\bar{t^1}_{L}t^1_{L}}/g_{s}(\tilde{k})= -5.18$ and $g_{G^1\bar{t^1}_{R}t^1_{R}}/g_{s}(\tilde{k})= -2.77$.}
        \label{collfig15}

\end{figure}

Searches for excited states of the top quark, decaying into third
generation fermions and either Higgs or gauge bosons have been the
subject of intensive recent study in the literature. Assuming QCD
production of these states, it has been shown that masses up to
about 1~TeV may be tested at the
LHC~\cite{Aguilar:2005},\cite{Holdom:2007},\cite{Witold:2007}. This
range of masses, however, falls short of the necessary one to test
the models under study, since the $t^1$-mass tends to be heavier
than 1~TeV. In this work, we will make use of the increased
production cross section of $t^1$ pairs induced by the presence of
$G^1$'s.  As shown in~\cite{Medina:2007hz}, there is a strong
correlation between the masses of the $G^1$ and the $t^1$ states,
$m_{G^1}\approx (2.2$~--~$3)\; m_{t^1}$. For the $G^1$-masses
allowed by precision measurements, $m_{G^1}\gtrsim 3$ TeV, the
$m_{G^1}$--$m_{t^1}$ correlation leads to $t^1$ masses above $1$
TeV. In figure~\ref{collfig15}, we show the different contributions
to the $t^1$ pair production cross-sections for a wide range of
$t^1$ masses beyond those allowed in our model. We fix $m_{G^1}=4$
TeV which corresponds in our framework to $m_{t^1}\sim 1.3$~--~$1.8$
TeV. It is precisely in this region of $t^1$ masses where the $G^1$
induced $t^1$ pair production cross-section becomes larger than the
QCD one. We notice in this figure that QCD production and $G^1$
induced production amplitudes interfere constructively. This can be
understood, by noticing that, contrary to the QCD case, the coupling
of the first excited KK gluon to $t^1$ pairs comes with an opposite
sign to the coupling of $G^1$ to first and second generation quarks.
In addition, most $t^1$ pairs are produced with an invariant mass
smaller than $m_{G^1}$, leading to a constructive interference
between both amplitudes due to the behavior of the gluon and $G^1$
propagators. .

With this in mind, we are going to show that it is possible to find a signal for $t^1$ with a mass up to about $1.5$~TeV when we include the
contribution to its production rate of the decay of a $G^1$ state with a mass of about $4$ TeV in the high luminosity LHC era. We also consider a
second point of 5D parameter space where the $G^1$-mass is of the order of 3.5 TeV and the $t^1$-mass is of order 1.25 TeV. We concentrate in a
region of parameter space with $c_{1}\geq 0$ where possible flavor changing operators are suppressed~\cite{Agashe:2006wa}. Moreover, as shown in
figures~\ref{collfig1} and~\ref{collfig12}, the suppressed couplings $g_{G^1\bar{t}_Lt_L}\sim g_{G^1\bar{t}_Rt_R}$ with respect to
$g_{G^1\bar{t^1}_Lt^1_L}$ lead to a large branching ratio for $t^1$ production induced by $G^1$'s. This can be appreciated in
figure~\ref{collfig14p}.
%

\subsection{Signal and Background simulation}

We are interested in the study of a mostly $SU(2)_L$ singlet, vector-like fermion state $t^1$, associated with the first KK resonance of charge
2/3.  This fermion state, due to gauge invariance, couples with regular SM QCD gluons with a strength given by the strong coupling and as
stressed before, its signatures have been studied in detail in~\cite{Aguilar:2005},\cite{Holdom:2007},\cite{Witold:2007}, showing a potential
reach for these particles with a mass up to about $m_{t^1}\sim 1$ TeV.

As explained in the previous section, in the Gauge-Higgs
unification models we are considering,
the $G^1$  decays mostly to $t^1$'s, and  $t^1$
prefers to decay  mainly in the $W^+b$ channel. Thus,
one of the  best options to study this resonance is by means
of the lepton+jets final state channel.

\begin{equation}
 pp \rightarrow (g+G^1) \rightarrow t^1\bar{t^1} \rightarrow W^{+}bW^{-}\bar{b}
                         \rightarrow l^{-}\bar{\nu}b\bar{b}jj,
\hspace{4mm} (l = e, \mu)
\end{equation}

The jets include quarks of the first two families and the charge conjugated process is also considered. The irreducible backgrounds for this
signal are coming from the SM by means of $t\bar{t}$ production and $W/Z + jets$, and from the exotic $t\bar{t}$ production induced by $G^1$. As
has been shown in ~\cite{Aguilar:2005},\cite{Holdom:2007}, the $W/Z +jets$ background can be reduced to negligible levels by requiring two
$b$-tags and lepton+MET. In this work, we impose the above requirements and therefore concentrate on the analysis of the main background from
$t\bar{t}$ production coming from both QCD and $G^1$.

We simulate the events for the background and signal at partonic level using the
Madgraph-Madevent package v4.1.33~\cite{Maltoni:2003}, which
includes the HELAS~\cite{Murayama:1992} subroutine to compute the matrix elements.
The renormalization scales for the QCD background are set at
the top mass, and for the backgrounds $G^1\rightarrow t\bar{t}$ and signal
$G^1\rightarrow t^1\bar{t^1}$  are set at the $t^1$-mass. Moreover, we
choose the set CTEQ5l~\cite{Lai:2000} for the structure functions. In order to reduce
the amount of events that are produced in the background
simulation due to the large QCD cross section, we put a mild cut at the generator
level for the $p_{t}$ of the lepton. This cut implies a weaker
restriction in comparison to the cuts that we will implement in the analysis and
therefore we do not introduce a bias over the background sample
by implementing it at the generator level. For the signal, we put no cuts at the
generator level and we produce almost ten times more events than
it is expected from the cross section. Therefore to obtain a realistic value
in the counting of events, it is necessary to normalize to the
correct value $\sigma \mathcal{L}$, where $\sigma$ is the corresponding cross section
and $\mathcal{L}$ is the LHC luminosity. This procedure
allows us to obtain smooth profiles and simulate the results at higher luminosities.

The partonic level events produced by Madgraph are passed through the Pythia~\cite{Sjostrand:2006} and PGS~4~\cite{Conway:2006} packages included
in  Madgraph. The Pythia package performs the hadronization process and includes initial and final state radiation, where at this stage we use
the default parameters that are included in Pythia. The PGS package performs a simulation of the LHC ATLAS detector based on a fit of the CDF
data. We then modify the default parameters for the $b$-tag efficiencies in order to have a more realistic simulation for the high luminosity LHC
era. To account for this, we replace the b-tag/mistag efficiencies with the set (1/2,1/10,1/30) for underlying $b$'s, $c$'s and untagged jets
(gluons and light quarks), respectively~\cite{Holdom:2007}. Furthermore, we set the jet cone reconstruction algorithm to $\Delta R = \sqrt{\Delta
\eta^{2} + \Delta \phi^{2}} = 0.6$ in order to optimize the reconstruction of the $W$ mass from very collimated
jets~\cite{Holdom:2007},\cite{Witold:2007}. This choice of $\Delta R$ enhances the ratio of signal over background due to the kinematic
differences between the $W$  plus b jet system arising from these two processes. We will continue with the discussion of this subject in the
analysis section, because it is found to be a key tool in the search for the $t^1$-resonance.

\begin{figure}[ht]

    \begin{minipage}[b]{0.5\linewidth}

       \centering
        \includegraphics[scale=0.45]{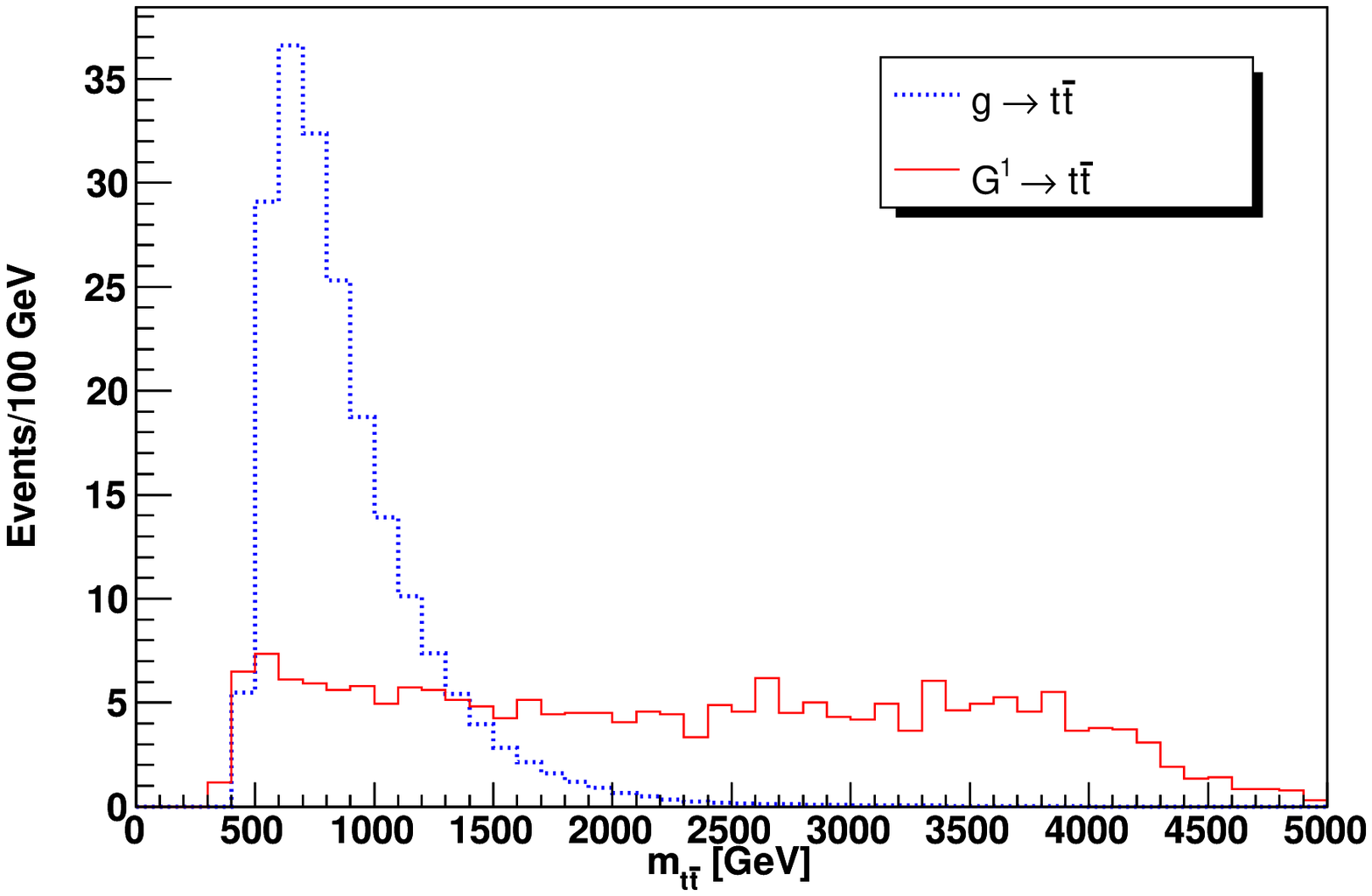}
        \caption{$t\bar{t}$ invariant mass reconstruction from QCD and $G^1$ induced production for point 1. Distributions normalized to 200 events.}
        \label{collfiginvtt}

    \end{minipage}
    \hspace{0.2cm}
    \begin{minipage}[b]{0.5\linewidth}

       \centering
        \includegraphics[scale=0.45]{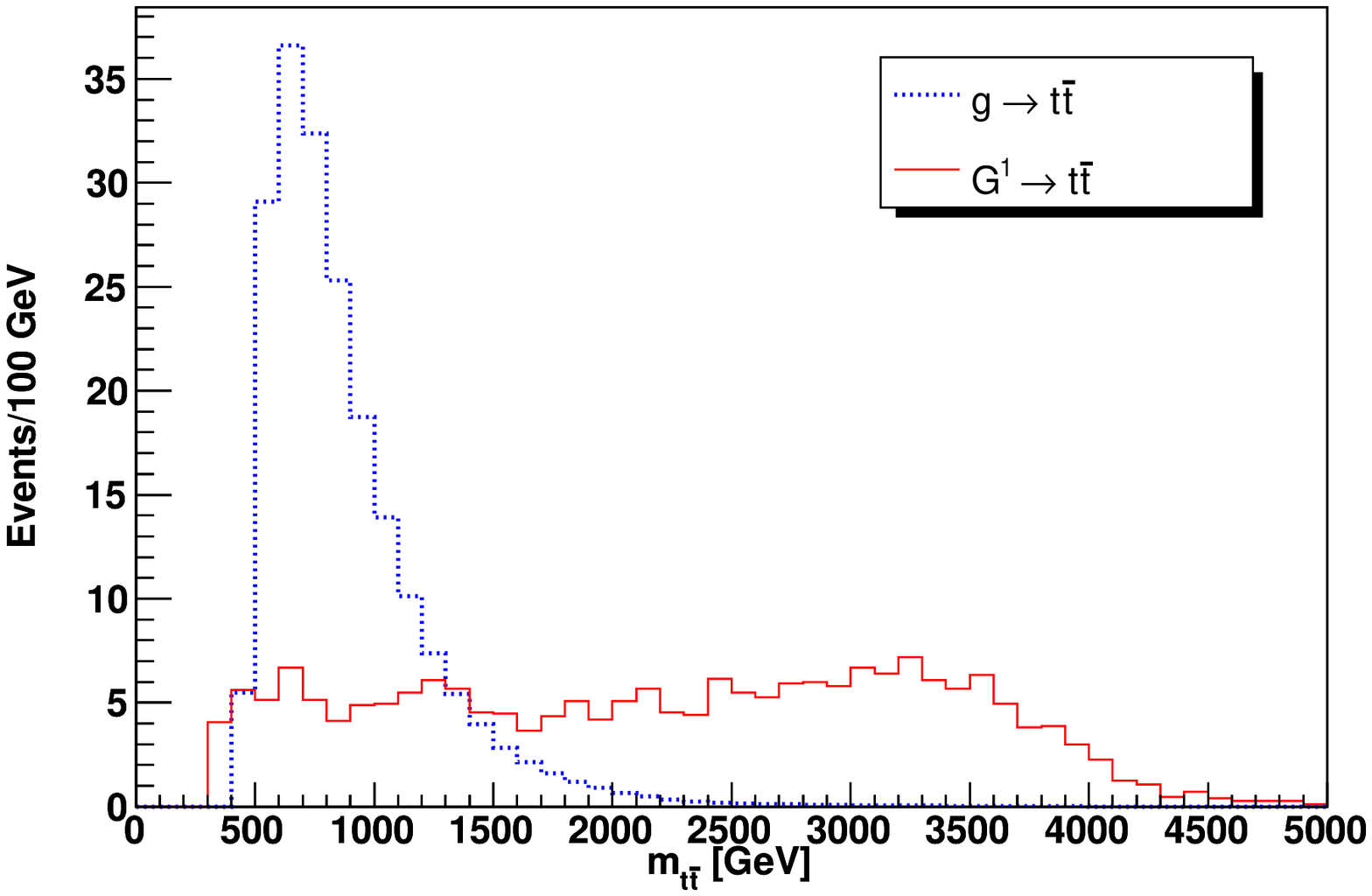}
        \caption{$t\bar{t}$ invariant mass reconstruction from QCD and $G^1$ induced production for point 2. Distributions normalized to 200 events.}
        \label{collfiginvtt2}
    \end{minipage}

\end{figure}

In table~\ref{point1} we specify the 5D parameters, associated
couplings and masses which we will use in the calculation of cross
sections and $t^1$-mass reconstruction. We will denote the
first set of parameters as point 1 and the second set as point 2.
\begin{table}[tbp]
\begin{center}
\begin{tabular}{|c|c|c|c|c|c|c|c|c|c|c|c|}
\hline
$c_{1}$ & $c_{2}$ & $c_{3}$ & $M_{B_{1}}$ & $M_{B_{2}}$ & $h/(\sqrt{2}f_h)$ & $m_{G^1}$ & $m_{t^1}$ & $g_{G^1\bar{t}t_R}$ & $g_{G^1\bar{t}t_L}$ &  $g_{G^1\bar{t^1}t^1_R}$ & $g_{G^1\bar{t^1}t^1_L}$ \\
\hline
0.26 & -0.41 & -0.57 & 2.2 & 0.4 & 0.278 & 3915.8 & 1470.2 &  -2.09 & -2.28 & -2.73 & -5.22 \\
\hline 0.24 & -0.41 &  -0.58 &  2.3 & 0.5 & 0.318 & 3439.6 & 1250.5 & -2.12 &   -2.50  &  -2.67  &  -5.20\\
\hline
\end{tabular}
\end{center}
\caption{Points of parameter space chosen for $t^1$ detection. All masses are given in GeV and the couplings are in units of $g_{s}(\tilde{k})$.}
\label{point1}
\end{table}
In table~\ref{colltable1} we show the values for the cross
sections and number of events expected at the parton level for
the different processes. One interesting aspect of this study as
can be seen in table~\ref{colltable1} is the relatively large
production cross section of top-quark pairs compaired to $t^1$
pairs from $G^1$-induced production. That this occurs in spite of
the fact that $\Gamma_{G^1tt}<\Gamma_{G^1t^1t^1}$ is a result of
the relatively large width of the $G^1$ particle that leads to a
strong departure from the narrow width approximation. Furthermore,
enhancements of PDF's at low x add to the width effects leading to a
preference in the production of tops proceeding from $G^1$'s at
relatively low $p_{t}$. The invariant mass of most of the
top-quark pairs produced from the $G^1$ decays is, therefore, much
smaller than the $G^1$ mass. This can be seen clearly in
figures~\ref{collfiginvtt} and ~\ref{collfiginvtt2}, where both
distributions have been normalized to a total of 200 events.

%


At this point we have not yet included  the contribution of K-factors to the cross section. These K-factors take into account NLO effects in the
computation of the matrix elements, and their direct effect is to enhance the value of the cross section for all processes. In~\cite{Holdom:2007}
it was shown that the signal to background ratio, for a heavy quark mass of $\sim 600$ GeV, computed with and without these K-factors is very
similar. Moreover, in the computation of the K-factors obtained in Ref.~\cite{Aguilar:2005}, the enhancement of the NLO cross section with
respect to the LO one increases with the mass of the heavy quark. It was obtain in Ref.~\cite{Aguilar:2005} that $K(\bar{t^1}t^1)\backsimeq 1.5$
and $K(\bar{t}t)\backsimeq 1.3$ for $m_{t^1}=1$~TeV. Thus, we expect that the inclusion of these K-factors will enhance our amount of events and
signal to background ratio. Even in the case in which the signal to background ratio remains approximately constant with the inclusion of
K-factors, there is still an increase in the significance of about  $\sqrt{K}$ that we will take into account at the end of our analysis.

\subsection{Event Selection}

The first selection of the events is done at the level of hadronized particles. In order to study the
properties of signal and background with sufficient statistics, we require that the hadronized
events fulfill the following two criteria:

\begin{itemize}
\item Isolated lepton with $p_{t} > 20$ GeV and $|\eta| < 2.5$ plus missing
      energy with $p_{t} > 20$ GeV.

\item At least three jets with $p_{t} > 20$ GeV and $|\eta| < 2.5$, with exactly two
      bottom-tags.
\end{itemize}

The lepton+MET requirements are useful to reduce the backgrounds from QCD jets, though we
will put a stronger cut on the lepton $p_{t}$ in the
final analysis. The condition of exactly two b-tags is useful to reduce the backgrounds
$W/Z + jets$ although it leads to a strong
reduction of the signal statistics.

\begin{table}[tbp]
\begin{center}
\begin{tabular}{|c|c|c|c|c|c|c|}
\hline
&\multicolumn{3}{c|}{Point 1} & \multicolumn{3}{c|}{Point 2}\\
\hline
Process &$\sigma$ [$fb$] & $N^{0}$ Events & $N^{0}$ after cuts & $\sigma$ [$fb$] & $N^{0}$ Events & $N^{0}$ after cuts\\
\hline\hline $G^1 \rightarrow t\bar{t}$& 4.12 &  1236 & 1 & 4.43 & 443 & 0 \\
$g \rightarrow t^1\bar{t^1}$ & 0.23 & 70 & 5 & 0.687 & 69 & 5 \\
$g + G^1 \rightarrow t\bar{t}$& 3025 & 907527 & 6 & 3085 & 308509 & 6\\
$g + G^1 \rightarrow t^1\bar{t^1}$ &0.88 & 266 & 21 & 2.015 & 201 & 13 \\
\hline
\end{tabular}
\end{center}
\caption{Value of the cross section at parton level before cuts for final state $l^{-}\bar{\nu}b\bar{b}jj$ including the charge conjugated
process, expected number of events, and number of events that pass the selection cuts in Eqs.~(\ref{selec_cuts1}) and~(\ref{selec_cuts2}), for
point 1 at 300 $fb^{-1}$ and point 2 at 100 $fb^{-1}$, respectively.} \label{colltable1}
\end{table}

\begin{figure}[tbp]

    \begin{minipage}[b]{0.5\linewidth}

        \centering
        \includegraphics[scale=0.48]{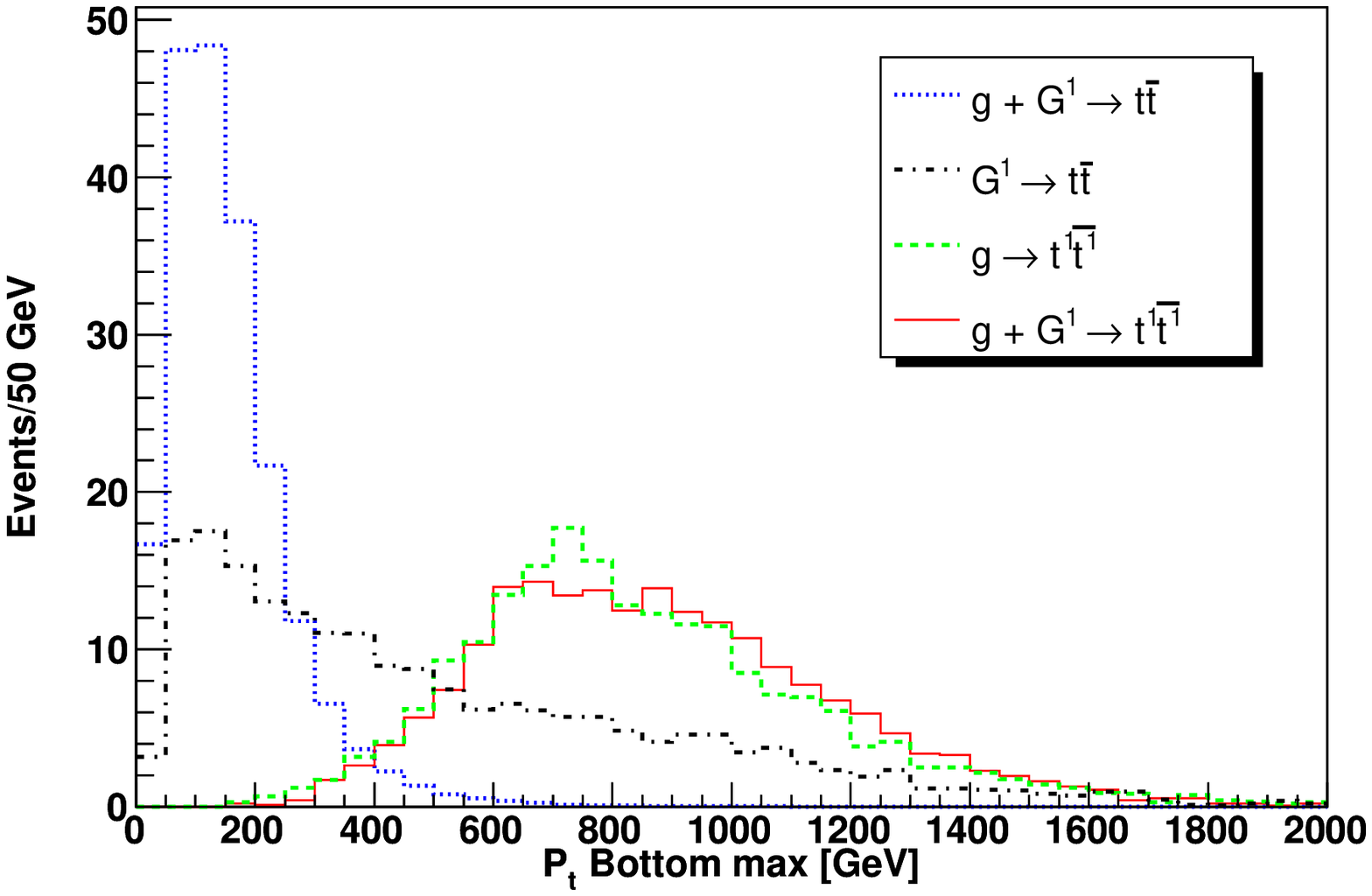}
        \caption{$p_{t}$ of the faster tagged bottom for point 1. Distribution normalized to 200 events.}
        \label{collfig23}

    \end{minipage}
    \hspace{0.2cm}
    \begin{minipage}[b]{0.5\linewidth}

        \centering
        \includegraphics[scale=0.48]{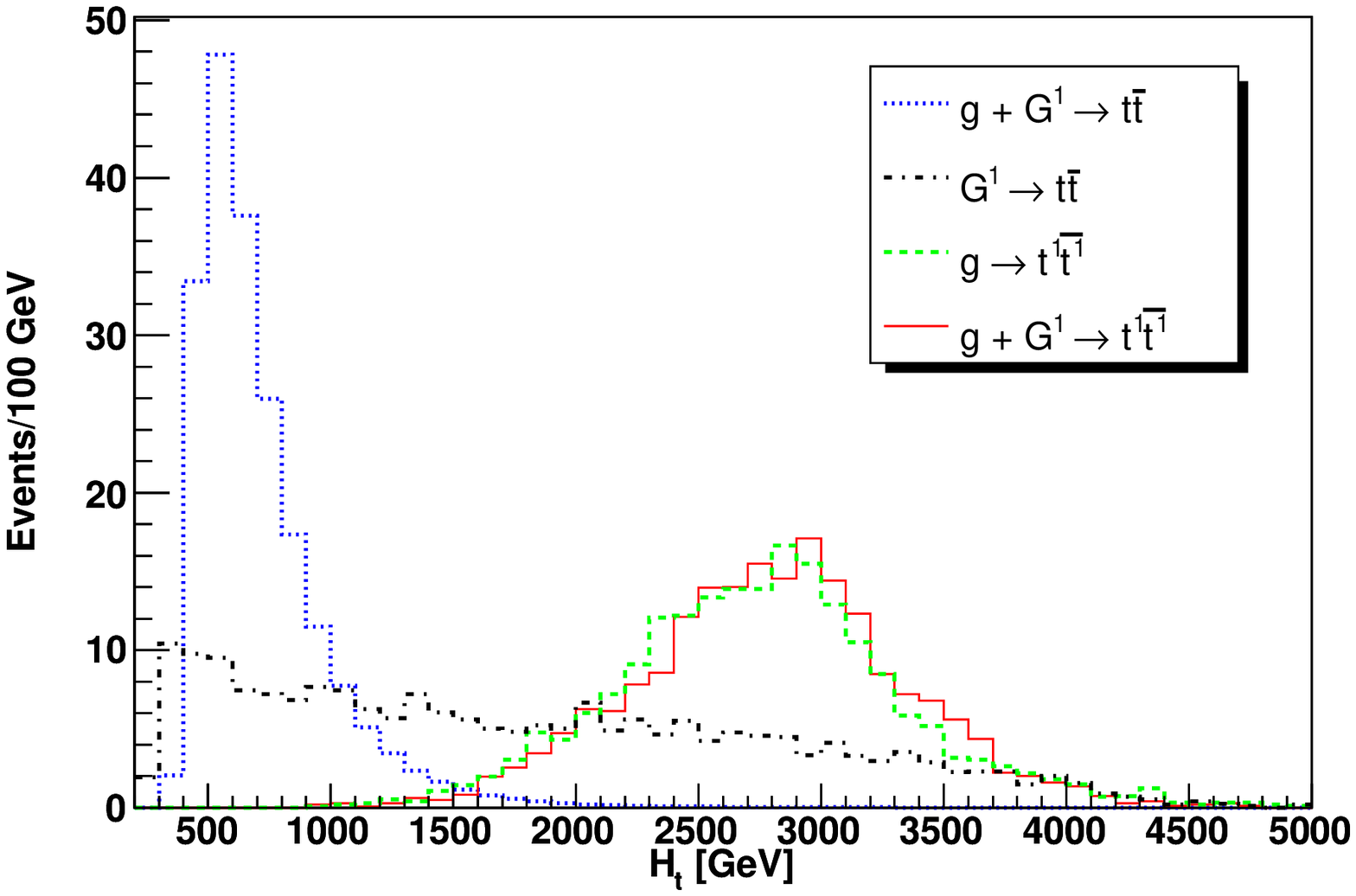}
        \caption{Scalar sum of the all $p_{t}$ for point 1. Distribution normalized to 200 events.}
        \label{collfig24}

    \end{minipage}

            \begin{minipage}[b]{0.5\linewidth}

        \centering
        \includegraphics[scale=0.48]{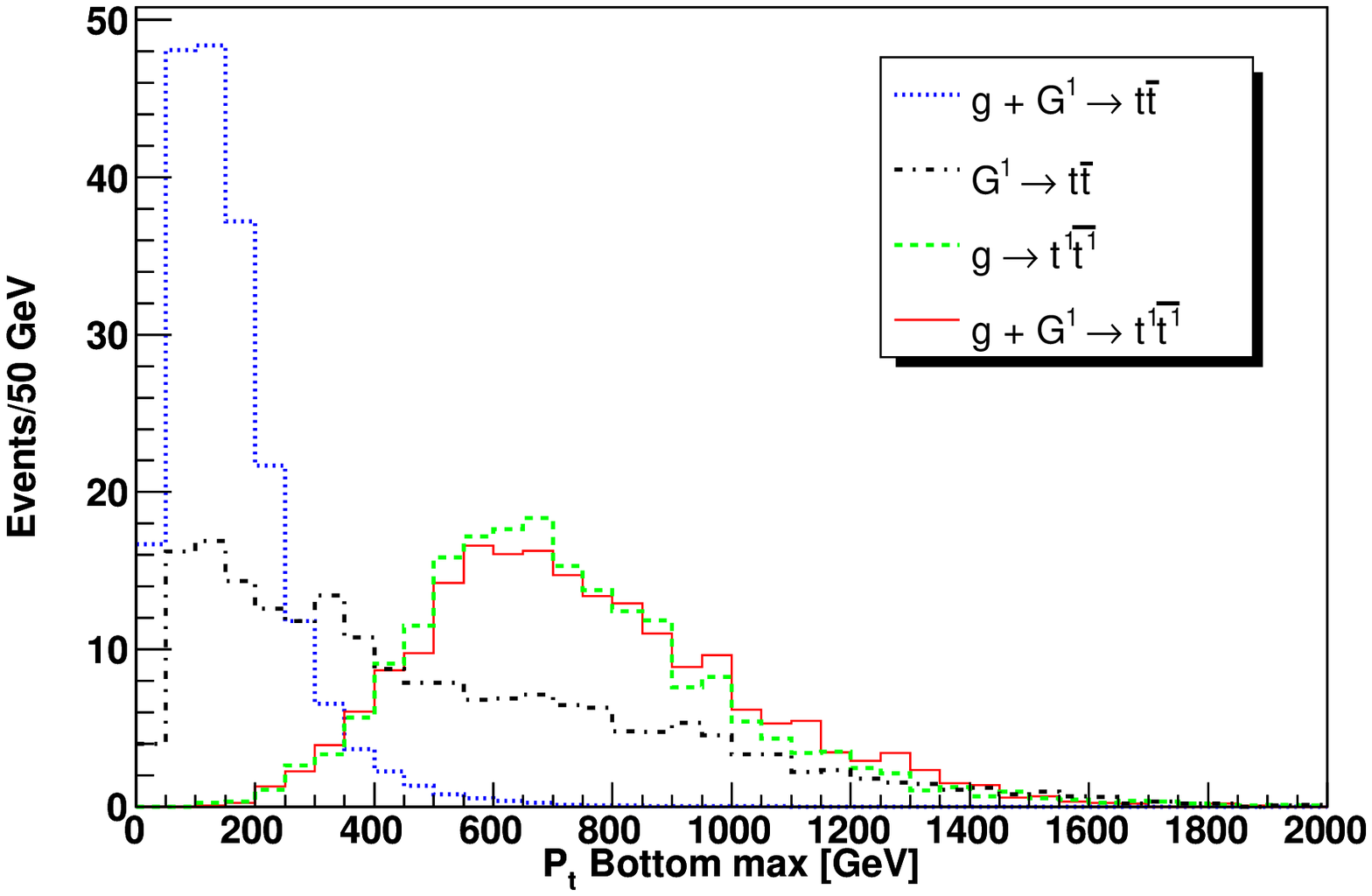}
        \caption{$p_{t}$ of the faster tagged bottom for point 2. Distribution normalized to 200 events.}
        \label{collfig25}

    \end{minipage}
    \hspace{0.2cm}
    \begin{minipage}[b]{0.5\linewidth}

        \centering
        \includegraphics[scale=0.48]{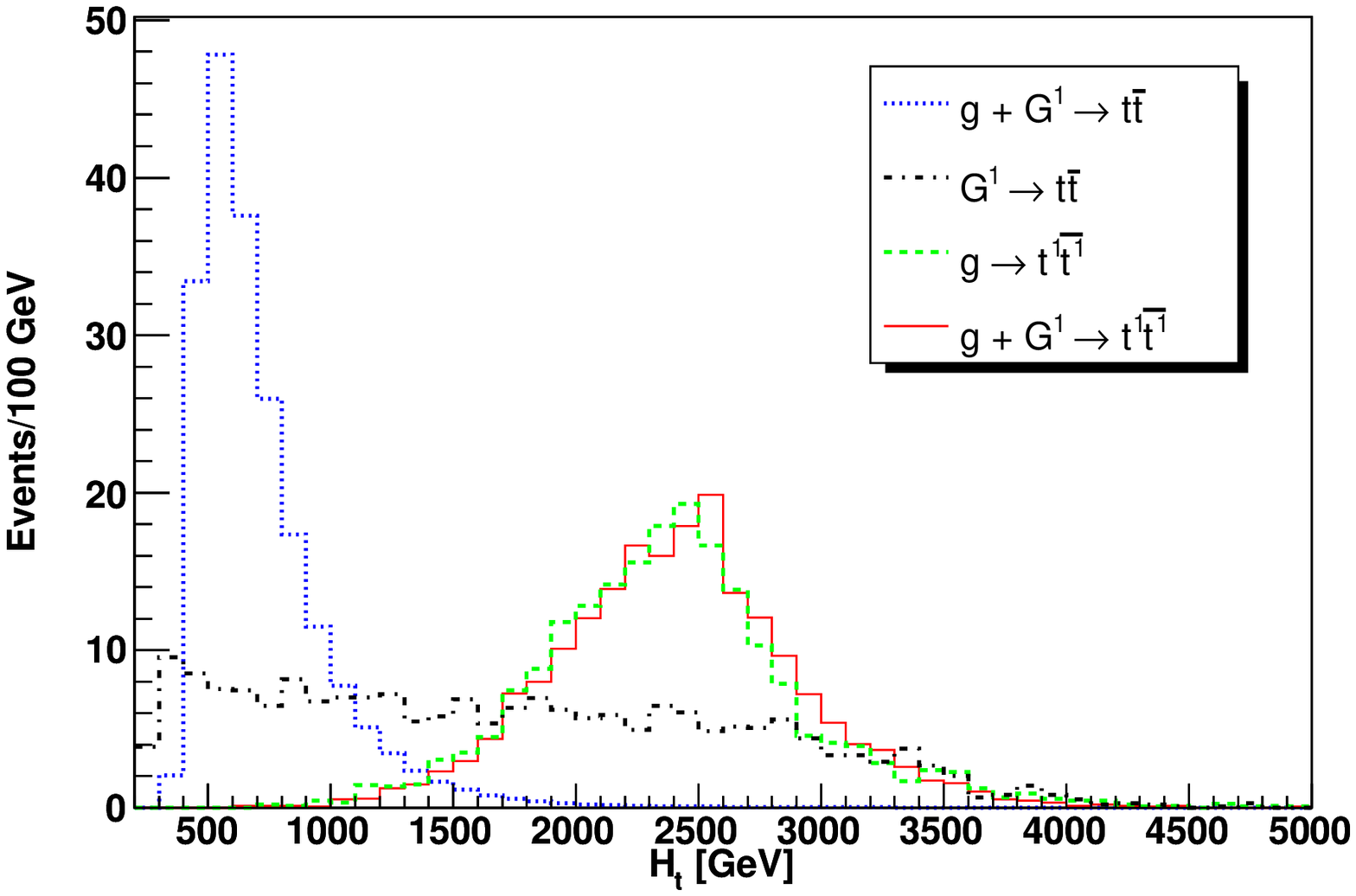}
        \caption{Scalar sum of the all $p_{t}$ for point 2. Distribution normalized to 200 events.}
        \label{collfig26}

    \end{minipage}

\end{figure}

>From the numbers in Table~\ref{colltable1}, it is clear that the cross section of the SM background is much bigger than the signal. Our approach
to reduce backgrounds to manageable levels follows the one in ~\cite{Holdom:2007}. We will put cuts over the $p_t$ of the harder bottom and over
$H_t$, the scalar sum of the $p_{t}$'s. The distribution of these two variables for the background and signal are shown in
figures~\ref{collfig23}--\ref{collfig26}. It is interesting to notice that the $p_{t}$ distribution of $t^1$ without the presence of $G^1$ is
similar to the same distribution including the $G^1$. From this behavior we conclude that, apart from the increase in the total production cross
section, the analysis of the $t^1$ signal in this scenario would not differ significantly from the previous analysis in the search of heavy
singlet quarks.
%

\subsection{Analysis}

To begin with, we would like to discuss the two different ways to reconstruct the $W$-mass from the hadronic jets which are used in the
literature to find this kind of resonances~\cite{Aguilar:2005},\cite{Holdom:2007},\cite{Holdom1:2007}. The first method is based in the direct
search of the two jets coming from the $W$ decay. The first jet is selected as the harder jet and the second one is the jet that forms an
invariant mass with the first jet close to the $W$-mass. Thus, to optimize this procedure it is necessary to use a cone reconstruction algorithm
with a $\Delta R \leqslant 0.4$ to avoid contamination from other jets. This method works very well when the events have $W$ jets which are not
too energetic, like in the $t\bar{t}$ background  and in the signal when the $t^1$ masses are much smaller than 1~TeV~\cite{Aguilar:2005}. The
second method consists of reconstructing the $W$-mass using the harder jet information. This method is based on the kinematic behavior of the
ultra-energetic jets coming from $W$'s. These jets, coming from highly boosted $W$'s, are so collimated that almost all of  the $W$-jet energy is
deposited in a cone of $\Delta R = 0.6$ around of the $E_{t}$ weighted barycenter~\cite{Holdom:2007}. In the scenario of heavy $t^1$'s, like the
ones present in our model, this method increases the signal and decreases the background. This is due to the fact that since the bottom produced
from the SM energetic top-quarks often contaminates the $W$-jets, then the reconstruction of the invariant $W$-mass is less efficient for the
background and the opposite effect works for the signal, where the bottom and $W$-jet are not that much collimated because the heavy $t^1$ is
produced almost at rest due to the broad $G^1$-width effect.

\begin{figure}[tbp]

    \begin{minipage}[b]{0.5\linewidth}

        \centering
        \includegraphics[scale=0.48]{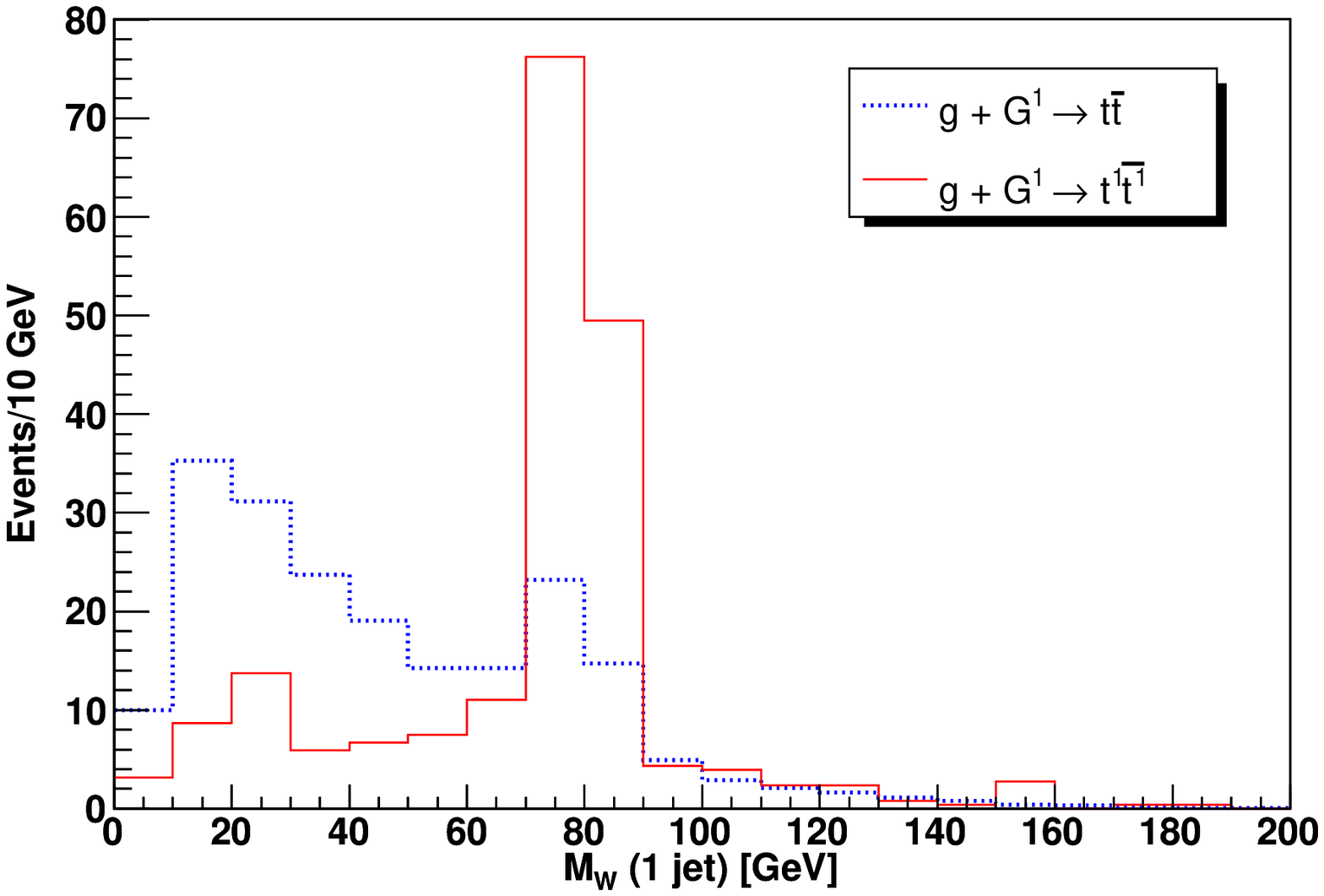}
        \caption{Invariant reconstructed W mass using the method of only one jet for point 1.
                 Distribution normalized to 200 events.}
        \label{collfig35}

    \end{minipage}
    \hspace{0.2cm}
    \begin{minipage}[b]{0.5\linewidth}

        \centering
        \includegraphics[scale=0.48]{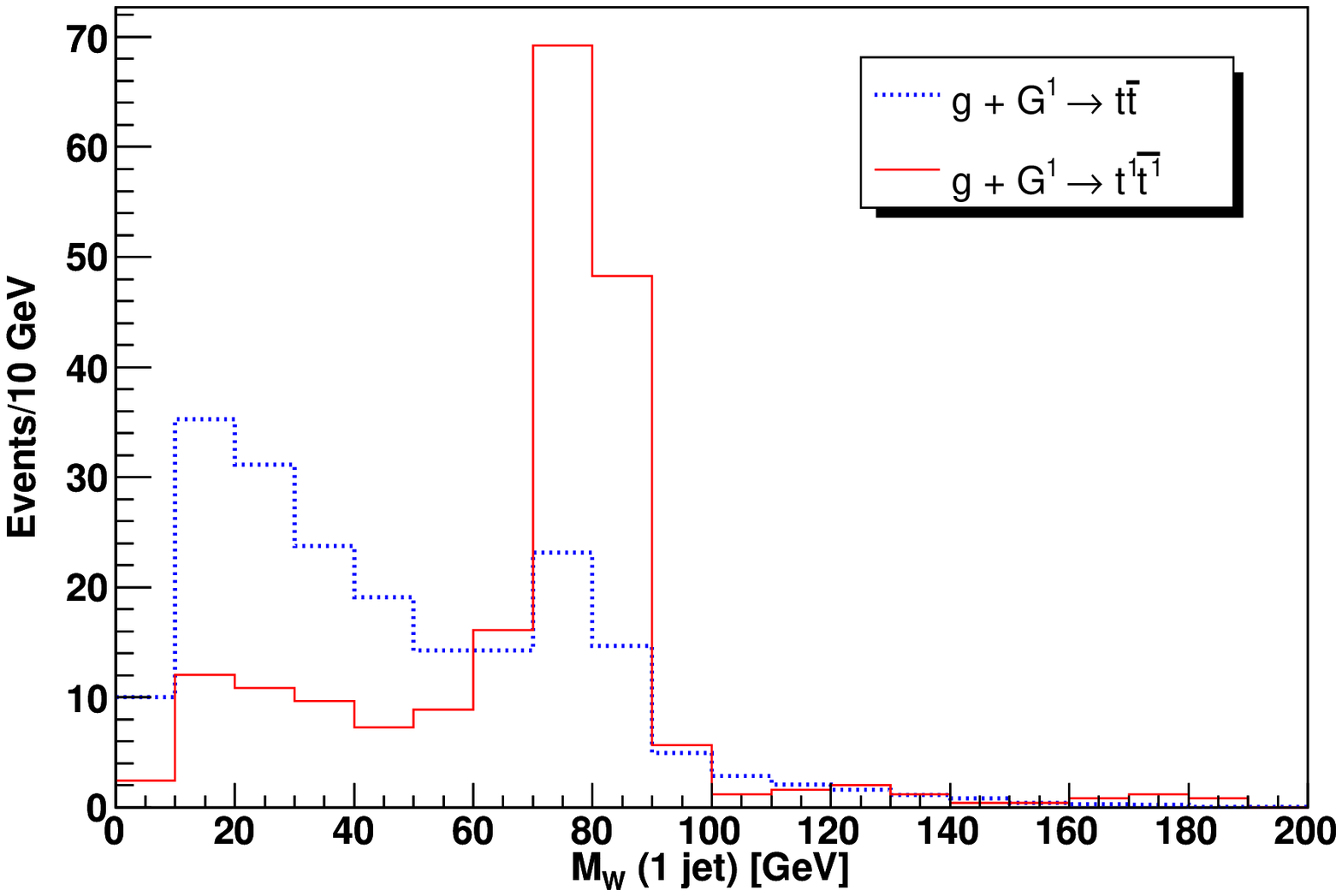}
        \caption{Invariant reconstructed W mass using the method of only one jet for point 2.
                 Distribution normalized to 200 events.}
        \label{collfig36}

    \end{minipage}

\centering
        \includegraphics[scale=0.48]{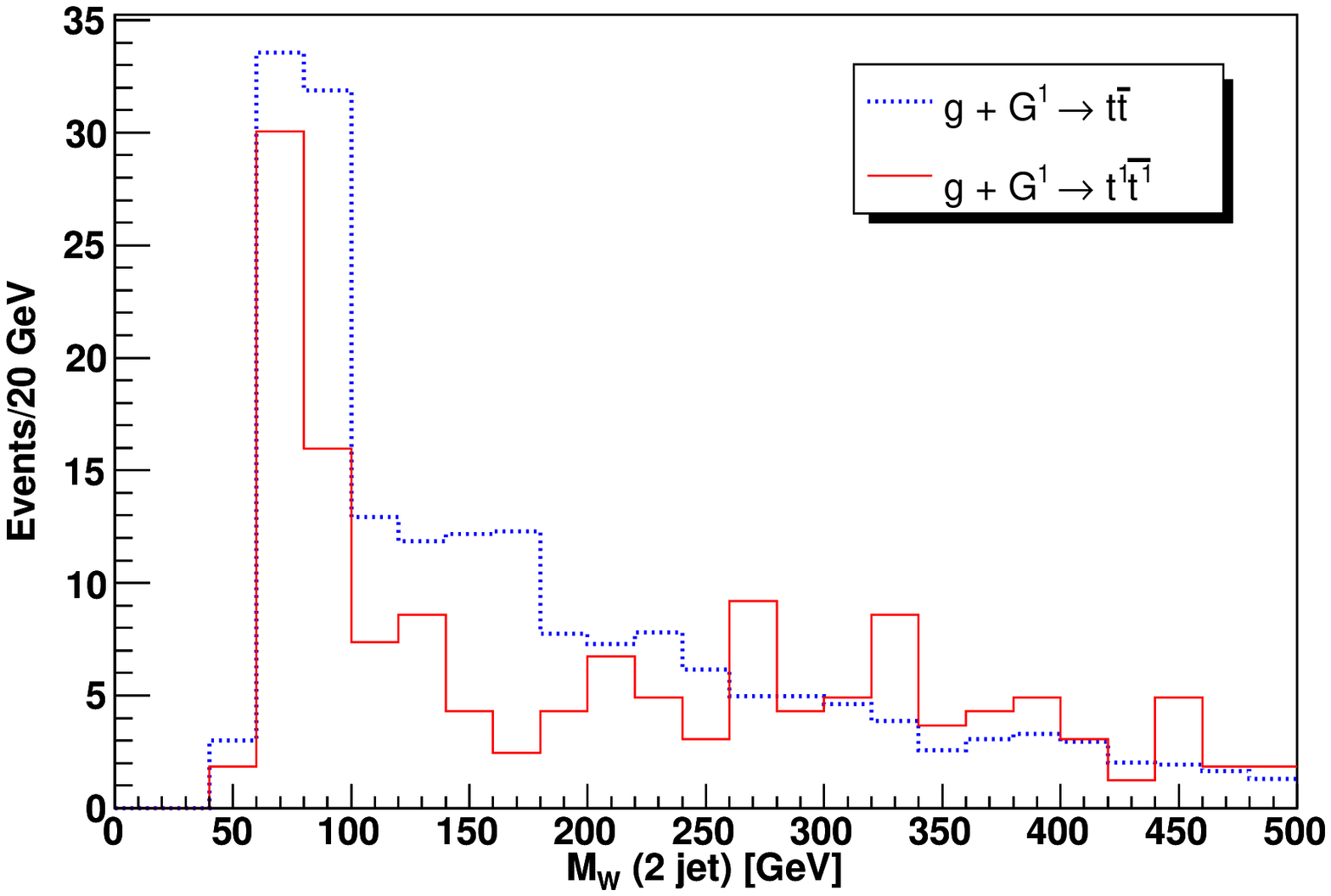}
        \caption{Invariant reconstructed W mass using the method of two jets, for point 1 (similar results for point 2).
                 Distribution normalized to 200 events.}
        \label{collfig37}

\end{figure}

Figures ~\ref{collfig35}--\ref{collfig37} show the distribution of the reconstructed invariant $W$-mass using both methods; the plotted events
are required to pass the first selection after Pythia-PGS level and $p_{t}^{j,max} \geqslant 150$ GeV. To make the comparison more transparent we
use a cone algorithm with $\Delta R = 0.4$ for the first method and $\Delta R = 0.6$ for the second one. From these histograms it is clear that
the more efficient method to reconstruct the invariant mass of the $W$ coming from $t^1$ is the one that treats the $W$ as a massive jet.
Furthermore, we can use this behavior in the invariant mass of the faster jet to strongly reduce the background from $t\bar{t}$.

Using the information from the $p_{t}$ and the invariant $W$-mass
distribution for signal and background, we select the next set of
cuts to do the reconstruction of the $t^1$ mass distribution. In the
case of $p_{t}^{b,max}$ and $H_{t}$ we use different cuts for point
1 (left column) and point 2 (right column), respectively,
\begin{eqnarray}
p_{t}^{b,max} &\geqslant& 350 \hspace{2mm} \text{GeV}, \quad\quad\quad\quad p_{t}^{b,max} \geqslant 300 \hspace{2mm} \text{GeV},   \nonumber \\
H_{t}    &\geqslant& 1900 \hspace{2mm} \text{GeV}, \quad\quad\quad\quad H_{t}\quad    \geqslant 1800 \hspace{2mm} \text{GeV},  \nonumber \\
p_{t}^{lepton} &\geqslant& 200 \hspace{2mm} \text{GeV},\quad\quad\quad\quad p_{t}^{lepton} \geqslant 150 \hspace{2mm} \text{GeV}.   \nonumber \\ \nonumber \\
\label{selec_cuts1}
\end{eqnarray}
The remaining cuts are common to both points,
\begin{eqnarray}
p_{t}^{j,max} &\geqslant& 250 \hspace{2mm} \text{GeV}, \nonumber \\
|M_{W} - M_{W}^{j}| &\leqslant& 20 \hspace{2mm} \text{GeV},\nonumber \\
|m_{Wb_{i}} - m_t| &\geqslant& 50 \hspace{2mm} \text{GeV}, \label{selec_cuts2}
\end{eqnarray}
where $m_{Wb_{i}}$ is the invariant mass of the $Wb$ system and by $b_{i}$ with $i=1,2$ we denote the two bottom-quarks which come from the decay
of either the $t^1$ or top pair system. As was mentioned before, the cut in the lepton $p_{t}$ is imposed in order to reduce the $\bar{t}t$
background, but leads to a strong reduction of the signal. The cut in $p_{t}^{j,max}$ strongly reduces the background while leaving the signal
after the first cut almost intact.  The last cut strongly reduces the background as well. This can be understood since in the $t$--$\bar{t}$
background case, at  least one of the bottoms tends to reconstruct the top-quark mass with the $W$ gauge boson; something that doesn't happen
with the signal. The relevance of the other cuts has been discussed above.

\begin{figure}[tbp]

  \begin{minipage}[b]{0.5\linewidth}

       \centering
        \includegraphics[scale=0.48]{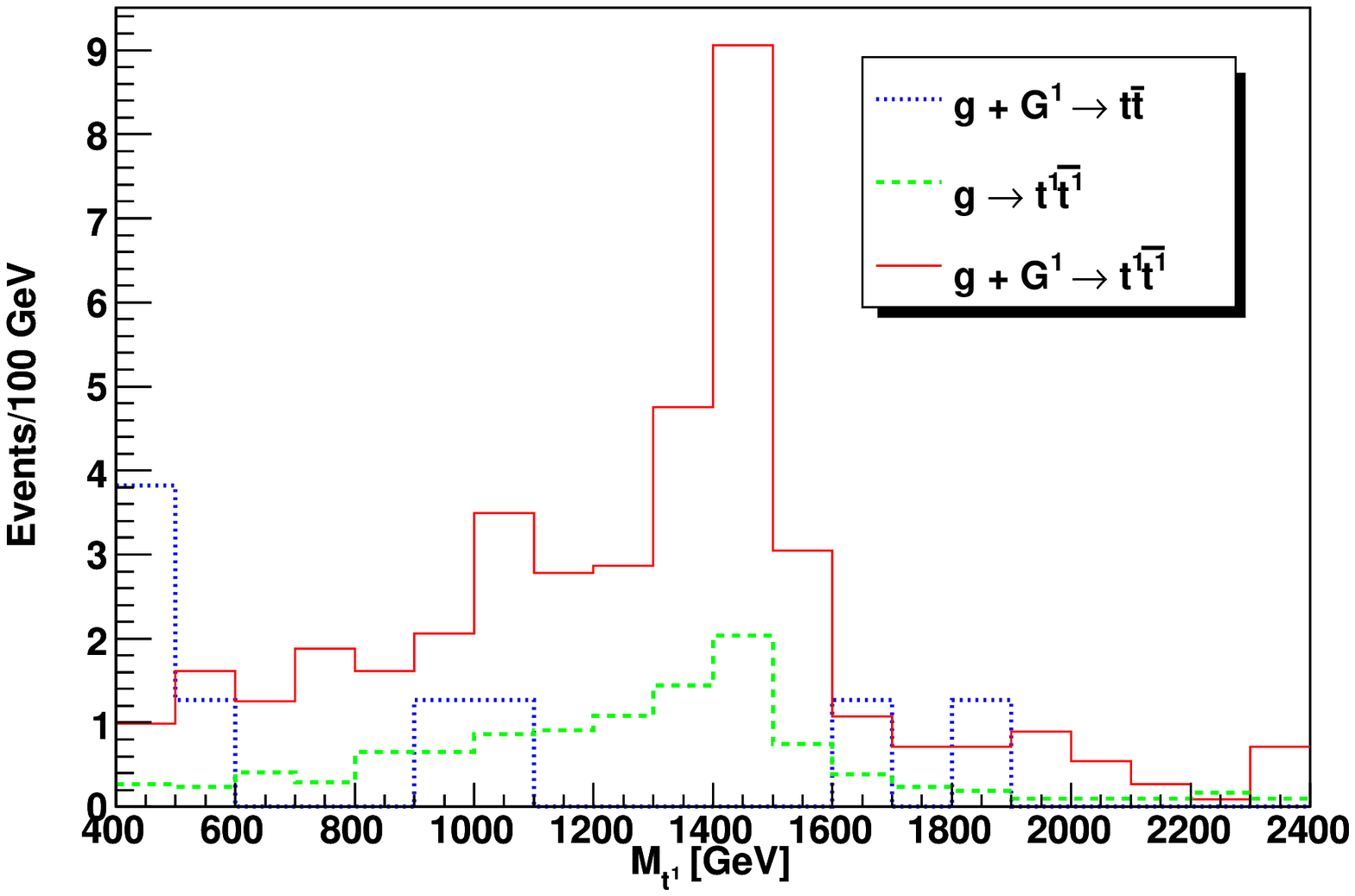}
        \caption{Distribution of the $t^1$ mass reconstructed hadronically for point 1. The numbers of events represent the results after 300 $fb^{-1}$ of LHC luminosity.}
\label{collfig57}

    \end{minipage}
    \hspace{0.2cm}
    \begin{minipage}[b]{0.5\linewidth}

        \centering
        \includegraphics[scale=0.48]{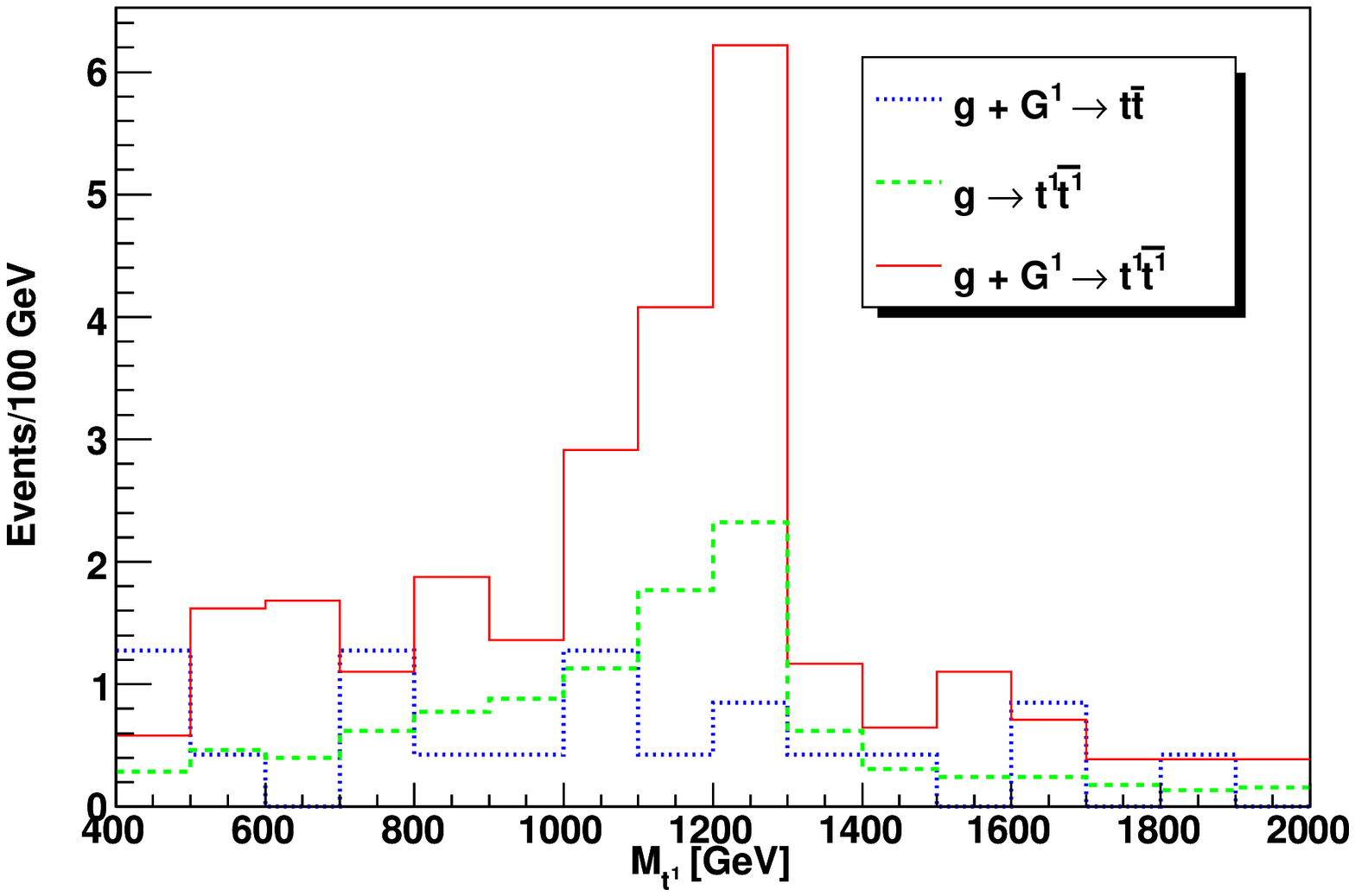}
        \caption{Distribution of the $t^1$ mass reconstructed hadronically for point 2. The numbers of events represent the results after 100 $fb^{-1}$ of LHC luminosity.}
\label{collfig58}
 \end{minipage}

\end{figure}

We present in figures \ref{collfig57} and \ref{collfig58} the reconstructed invariant
mass $m_{t^1}$ vs. the number of reconstructed $t^1$'s per 100 GeV for both signal and background
at 300~$fb^{-1}$ of LHC luminosity for point 1 and at 100~$fb^{-1}$ for point 2. In both cases we notice
an accumulation of events with an invariant mass close to the physical mass of
the $t^1$ particles,  leading to the first hint of a heavy quark with
a relevant branching ratio decay in the $W^{+}b$ channel. It is apparent from both figures that
the background has a much flatter distribution
in the same mass range after cuts. Following Ref.~\cite{Holdom:2007} we have considered
all $W$~$b$ pairs in each event. This leads to an increase in statistics by doubling the
number of points per event and furthermore a broadening of the reconstructed mass distribution.

\begin{figure}[tbp]

  \begin{minipage}[b]{0.5\linewidth}

       \centering
        \includegraphics[scale=0.48]{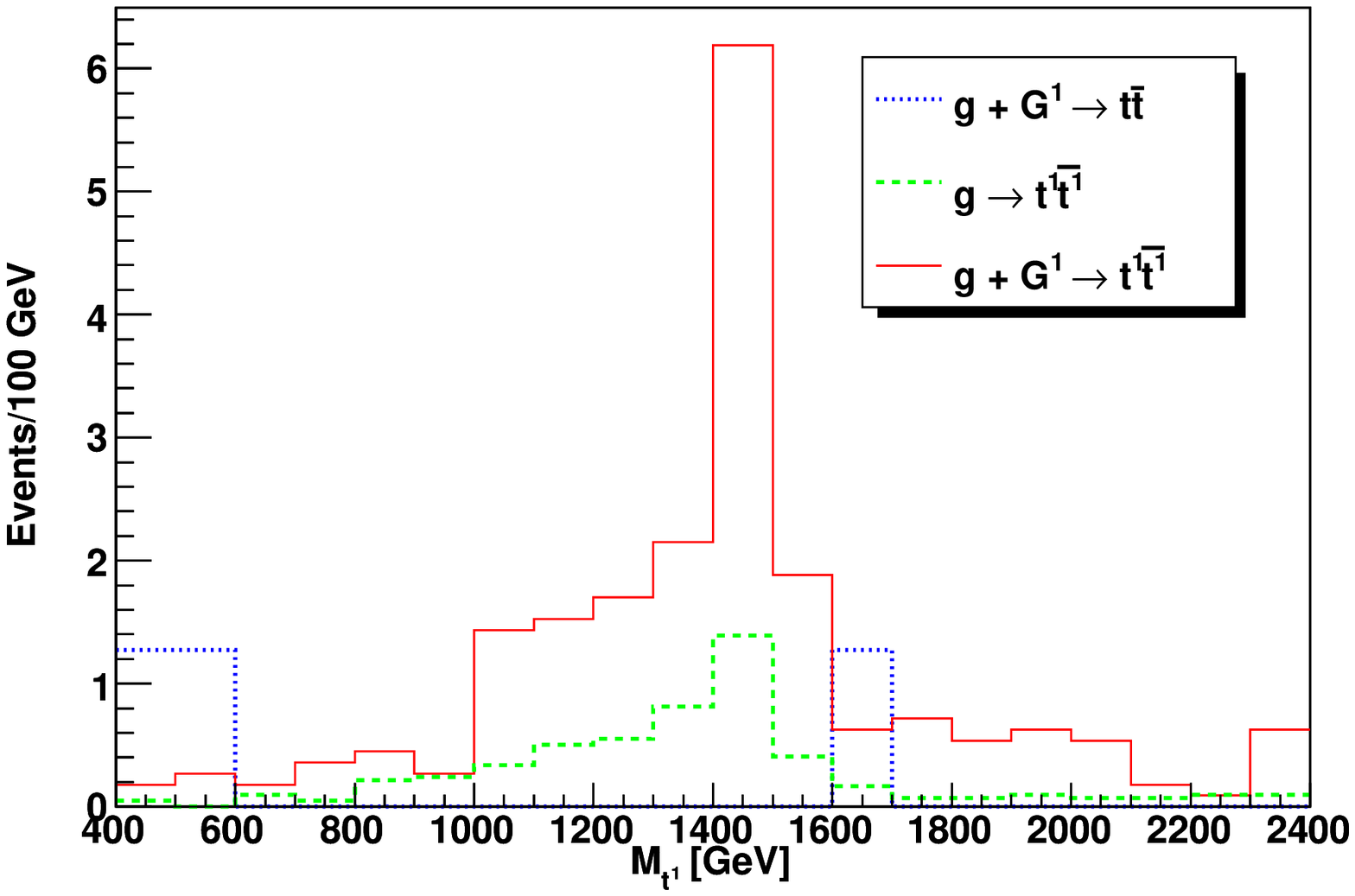}
        \caption{Distribution of the $t^1$ mass reconstructed hadronically for point 1. The numbers of events represent the results after 300 $fb^{-1}$ of LHC luminosity,
        with the additional requirement of $\Delta R$ in the Wb system.}
        \label{collfig47}

    \end{minipage}
    \hspace{0.2cm}
    \begin{minipage}[b]{0.5\linewidth}

        \centering
        \includegraphics[scale=0.48]{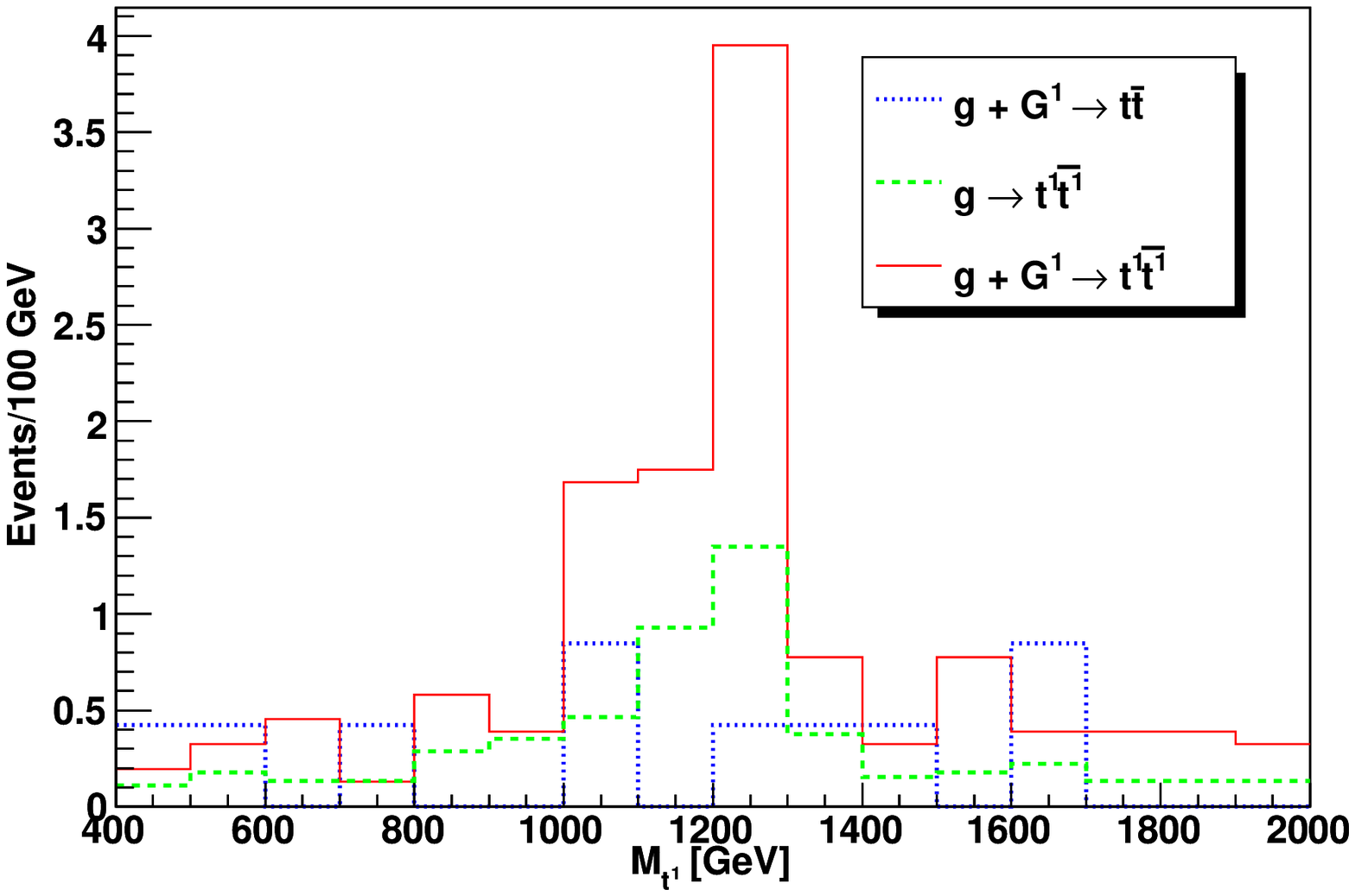}
        \caption{Distribution of the $t^1$ mass reconstructed hadronically for point 2. The numbers of events represent the results after 100 $fb^{-1}$ of LHC luminosity,
        with the additional requirement of $\Delta R$ in the Wb system.}
        \label{col48}
 \end{minipage}

\end{figure}

In order to distinguish among the bottom-quarks which we use to reconstruct the $t^1$ resonance, we can impose a further requirement. We will
choose the bottom which has the biggest $\Delta R$ with respect to the $W$ gauge boson. This requirement is imposed because the bottom quark and
the $W$'s coming from $t^1$ decays usually have a large angular separation since $t^1$'s are produced mostly at low $p_t$. This will lead to a
preference of real  reconstructed $t^1$'s compared to fake ones in the signal events. We plot in figures~\ref{collfig47} and~\ref{col48} the
reconstructed invariant mass $m_{t^1}$ vs. the number of reconstructed $t^1$'s per 100 GeV for both signal and background at 300~$fb^{-1}$ of LHC
luminosity for point 1 and at 100~$fb^{-1}$ for point 2, with this new requirement.

In order to estimate the statistical significance of the excess of
signal over background for points 1 and 2 at the respective
luminosities, we take into account the total number of events of
signal (S) and background (B) obtained after cuts. Assuming a
Poisson distribution of signal and background events, the
statistical significance would be given by $S/\sqrt{S+B}$. Using
the numbers in Table~\ref{colltable1} for points 1 and 2, we
estimate a statistical significance $S/\sqrt{S+B} \sim
4.04$~$\sigma$ at a luminosity of 300~$fb^{-1}$ and $S/\sqrt{S+B}
\sim 2.98$~$\sigma$ at a luminosity of 100~$fb^{-1}$,
respectively. As mentioned before, the inclusion of K-factors will
tend to increase the total significance of the signal. Taking
values of the K-factors of roughly 1.5 for both signal and
background, we obtain a 5~$\sigma$ significance for point 1 at
300~$fb^{-1}$. For point 2, instead, $t^1$-discovery cannot be
achieved even after the inclusion of K factors at a luminosity of
100~$fb^{-1}$. However, an increase in luminosity up to about
200~$fb^{-1}$ will be sufficient to discover $t^1$ at 5~$\sigma$
significance.

An improved analysis based on an exhaustive treatment of physics and non-physics backgrounds, with further optimization of the cuts, inclusion of
NLO QCD corrections, and a more realistic detector simulation should be done to confirm these results. Keeping this in mind, our preliminary
results indicate that, after including K-factors, 3~$\sigma$ evidence for the presence of these particles may be found already at 100~$fb^{-1}$
(60~$fb^{-1}$) and that 5~$\sigma$ discovery can be achieved with about 300~$fb^{-1}$ (200~$fb^{-1}$) for points 1 and 2, respectively.
Figures~\ref{collfig47} and~\ref{col48} clearly show the importance of the contribution of $G^1$'s in  $t^1$ production (and detection), by
comparing the number of events produced by the process $g+G^1\rightarrow\bar{t^1}t^1$ with the ones produced by the process $g\rightarrow
\bar{t^1}t^1$. If we only had direct QCD production the signal would be overwhelmed by the background.

In order to explore the last point further,  we have done a simulation considering only QCD production of $t^1$ pairs at the LHC with a
luminosity of 300~$fb^{-1}$. Assuming the biggest branching ratio into the $W^{+}b$ channel for a given $t^1$-mass as seen in figure 4,
($BR_{W^{+}b}\approx0.45$), using the massive jet technique and similar cuts and efficiencies as was done in the previous simulations, we found
that the maximum $t^1$-mass which can be reconstructed leading to a 5~$\sigma$ discovery at the LHC is $m_{t^1}\approx 1.1$~TeV. This corresponds
to a $t^1$ production cross section of about 30~fb. Similarly, for point 1 we have shown that by considering $G^1$-induced production of $t^1$ in
addition to QCD production at a luminosity of 300~$fb^{-1}$, the LHC will be sensitive to $t^1$ masses up to about 1.5 TeV. This corresponds to a
$t^1$ production cross section of about 15~fb as presented in figure~\ref{collfig15}. Though the cross section for point 1 is half the value of
the pure QCD cross section necessary to discover a 1.1 TeV $t^1$, the difference can be understood by the increase in the efficiency of the cuts
at high mass.

\begin{figure}[tbp]

        \centering
        \includegraphics[scale=0.6, angle=90]{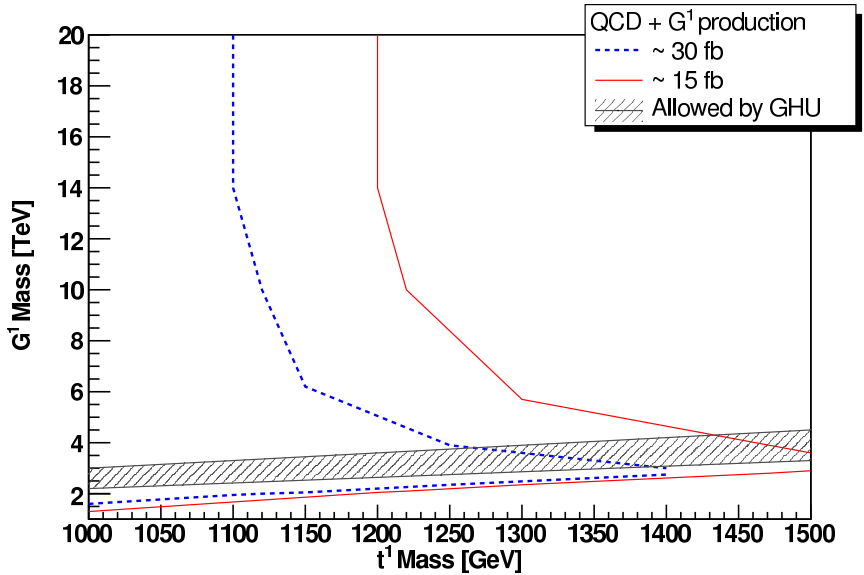}
        \caption{Curves of constant cross section for QCD in addition of $G^1$ decay, in $(m_{G^1},m_{t^1})$ plane.}
        \label{collfig49}

\end{figure}

We can use the above information to obtain an estimate of the reach of the LHC at 300~$fb^{-1}$ for arbitrary $m_{t^1}$ and $m_{G^1}$ masses. In
figure~\ref{collfig49} we show the curves representing the points in the $(m_{G^1},m_{t^1})$ plane with constant cross section of either 30~fb
(blue dotted line) or 15~fb (red solid line). The lower bounds in the figure correspond to phase suppression due to the impossibility of
producing $t^1$'s from $G^1$-decays. For masses $m_{G^1}\rightarrow +\infty$, the blue curve approaches the maximum value for $t^1$-masses that
can be tested by QCD production alone. All points between the upper and lower blue and red lines have a cross section larger than 30~fb and
15~fb, respectively. Since the efficiency of the cuts tends to increase with large $t^1$-masses, one expects that the LHC will be able to explore
the whole region bounded by the lower and upper blue curves with $m_{t^1} \geq 1.1$ TeV. This, however, underestimates the full reach of the LHC
since it doesn't take into account the increase in cut efficiency at large $t^1$ masses. The real sensitivity curve for the case of $G^1$-induced
production in addition to QCD production, will lie in between the upper red and blue curves, interpolating the points $(+\infty, 1.1\text{TeV})$
and $(4\text{TeV},1.5 \text{TeV})$ in $(m_{G^1},m_{t^1})$ space. The shaded area in Figure~\ref{collfig49} represents the region of allowed $G^1$
and $t^1$ masses in the Gauge-Higgs unification models considered in the previous sections. We can see from the figure that the LHC at
300~$fb^{-1}$ will be able to probe a large region of parameter space for these type of models which are consistent with electroweak symmetry
breaking and electroweak precision tests.

\subsection{Discussion and Outlook}

Let us close this section by discussing some important points related to the search for the $t^1$ resonance. In the analysis above, we have
considered a flat 0.5 b-tagging efficiency, independent of the $p_t$ of the tagged jet. This may be an optimistic assumption, particularly at the
early stages of the LHC.  The search for the heavy $t^1$'s analyzed in this work, however, will require high LHC luminosity. If the LHC b-tagging
efficiency at high $p_{t}$ for two b-tagged jets proves to be lower than the one assumed in this work, it may be helpful to perform the search
without b-tagging  requirements. This will increase both signal and background, particularly the $W$~plus jets background. The cut on the $p_t$
of the harder jet and the  massive $W$-jet, and the requirement of a high-$p_t$ lepton and large missing energy may be sufficient to reduce the
background to manageable levels. The results of Ref.~\cite{Holdom1:2007} seem promising in this direction.

In this work, we have only analyzed the search for the $t^1$-quark in its $W$ plus $b$ decay mode. However, as discussed above, $t^1$ decays
about half of the time into a top quark and either a Higgs or a Z-gauge boson. For the large $t^1$ masses  we considered in this work, the decay
products will be highly boosted and therefore the reconstruction of these particles will be most efficient by using the technique of  massive
jets. It would be useful to analyze the LHC reach in the search for $t^1$'s including these alternative decay modes.

Single $t^1$ production is a complimentary production channel to the
one discussed in this work and has been analyzed in the context of
the little Higgs Model \cite{Azuelos:2004dm,Han:2005ru}. Similar to
the $\bar{t}^1 t^1$ production induced by $G^1$, the single $t^1$
production cross-section also becomes larger than the QCD $t^1$ pair
production cross-section for sufficiently large values of the $t^1$
masses. The study of Ref.~\cite{Azuelos:2004dm,Han:2005ru} shows
that the $t^1$ single production extends the sensitivity to $t^1$
masses at the LHC well above those reachable via the QCD process. It
would be interesting to analyze single $t^1$ production in detail in
our specific model. In addition to likely extending the reach for
the $t^1$ mass, this could also provide information about the $W t^1
b$ coupling.

One important property of the Gauge-Higgs unification model analyzed in this article is the suppression of the Higgs production cross section in
both the $W$-fusion and gluon fusion modes~\cite{Carena:2007ua,Sakamura:2006rf,Medina:2007hz,Falkowski:2007hz,Djouadi:2007fm,Maru:2007xn}. The
reduction of the Higgs boson coupling to gluons could imply the presence of extra scalar states that mix with the Higgs and/or of extra colored
states. In the class of Gauge-Higgs unification models under analysis, the $t^1$ will provide the dominant contribution to the gluon fusion
amplitude in addition to the top quark. In this sense,  the discovery of the heavy $t^1$ mode, and the study of its properties, may be essential
in order to understand Higgs boson production at the LHC.

\section{Conclusions}

We have analyzed the possible detection at the LHC of excited states of the top quark in Gauge-Higgs Unification models in warped extra
dimensions that are consistent with radiative electroweak symmetry breaking and precision electroweak measurements. In these models, the $t^1$
tends to be strongly coupled to the first gluon excitation $G^1$ and sufficiently light, so that it can be produced from decays of $G^1$. This
reduces the branching ratio of the decay of $G^1$ into top quarks and enhances the width of $G^1$, making its reconstruction from top decays
difficult. On the other hand, consistency with precision measurements may only be obtained for masses of $t^1$ larger than about 1~TeV.

After presenting a consistent functional method for the computation of the couplings of the fermion modes to the gauge modes, we computed the
relevant decay widths necessary for the collider phenomenology analysis. An immediate consequence of the heaviness of the $t^1$ is that its decay
branching ratios into $W^+b$, $tZ$ and $tH$ states are in an approximate $2:1:1$ relation which becomes a better approximation with increasing
$t^1$-mass. On the other hand, for the large $t^1$ masses we considered, $m_{t^1}>1$ TeV, the SM decay products tend to be boosted, making the
reconstruction of the weak gauge bosons from its decay into quarks very difficult. Instead, as recently emphasized in the literature, it is
convenient to apply the technique of massive jets, that becomes essential for an efficient reconstruction of the $t^1$ states in our case.

Our analysis shows that the increase in the production cross section of  $t^1$ pairs, induced by the presence of $G^1$'s in the class of models
under analysis, extends the reach of the $t^1$ searches to masses not accessible via direct QCD production. There is a strong correlation between
the $G^1$ and $t^1$ masses, and we obtain that for a value of $m_{G^1}$ of about 4~TeV and with a high LHC luminosity of about 300 $fb^{-1}$, the
first KK excitation of the top quark with a mass of 1.5~TeV may be detected. The discovery of a $t^1$ in this mass range, possible due to an
increase in its production cross section, may point towards the presence of a heavy gauge boson resonance contributing to the $t^1$ production
rate. It may be possible to obtain information about the $G^1$ resonance from the analysis of $t^1$ production. Moreover, while $t^1$ induces a
suppression of the Higgs production via gluon fusion, the study of the additional decay channel of $t^1$ into Higgs and a top-quark may lead to
novel Higgs production signatures at the LHC. In summary, the search of a singlet vector-like heavy quark  pair produced at the LHC, opens the
possibility to explore the parameter space of Gauge-Higgs unification models consistent with precision electroweak measurements and electroweak
symmetry breaking.
\section*{Acknowledgements}

We would like to thank  Adam Falkowski, Ayres Freitas, Jay Hubisz, Tom LeCompte, Ben Lillie, Ian Low, Eduardo Ponton, Jose Santiago, Jing Shu and
Tim Tait for useful discussions and comments. Work at ANL is supported in part by the US DOE, Div.\ of HEP, Contract DE-AC02-06CH11357. This work
was also supported in part by the U.S. Department of Energy through Grant No. DE-FG02-90ER40560. B.P. work was supported in part by CONICYT,
Chile, and by a Fermilab Graduate Student Fellowship.

\newpage
\appendix
{\Large{{\bf APPENDIX}}}

\section{$SO(5)$ Generators}
\label{SO5}
\renewcommand{\theequation}{A.\arabic{equation}}
\setcounter{equation}{0}  

\renewcommand{\thefigure}{A.\arabic{figure}}
\setcounter{figure}{0}  
The generators of the fundamental representation of $SO(5)$, with
$\tr[T^\alpha.T^\beta]=C(5)\delta^{\alpha\beta}$, are given by:

\begin{eqnarray}
\label{gen} T^{a_{\rm
L,R}}_{i,j}&=&-\frac{i\sqrt{C(5)}}{2}\left[\frac{1}{2}\epsilon^{abc}(\delta_i^b\delta_j^c-\delta^b_j\delta^c_i)\pm(\delta^a_i\delta^4_j-\delta^a_j\delta^4_i)\right],\nonumber\\
T^\ha_{i,j}&=&-i\sqrt{\frac{C(5)}{2}}(\delta^\ha_i\delta^5_j-\delta^\ha_j\delta^5_i),
\end{eqnarray}
where $T^\ha$ $(\ha=1,2,3,4)$ and $T^{a_{\rm L,R}}$ $(a_{\rm
L,R}=1,2,3)$ are the generators of $SO(5)/SO(4)$ and $SO(4)$
respectively.

Define the matrices $A$ and $B$ as follows, which gives us the
generators in the basis that $T^{3_L}$ and $T^{3_R}$ are diagonal:

\begin{eqnarray}
\label{A.mat}
\begin{array}{ccccc}
 A &=& \frac{1}{\sqrt{2}}\begin{pmatrix}
-i & -1 & 0 & 0 & 0 \\
0 & 0 & -i & 1 & 0 \\
0 & 0 & i & 1 & 0 \\
-i & 1 & 0 & 0 & 0 \\
0 & 0 & 0 & 0 & \sqrt{2}
\end{pmatrix}
\end{array}
\end{eqnarray}

\begin{eqnarray}
\label{B.mat}
\begin{array}{cccccccccc}
B &=& \frac{1}{\sqrt{2}}\begin{pmatrix}
i & 1 & 0 & 0 & 0 & 0 & 0 & 0 & 0 & 0 \\
0 & 0 & -i & 1 & 0 & 0 & 0 & 0 & 0 & 0 \\
0 & 0 & -i & -1 & 0 & 0 & 0 & 0 & 0 & 0 \\
-i & 1 & 0 & 0 & 0 & 0 & 0 & 0 & 0 & 0 \\
0 & 0 & 0 & 0 & -1 & i & 0 & 0 & 0 & 0 \\
0 & 0 & 0 & 0 & 0 & 0 & \sqrt{2} & 0 & 0 & 0 \\
0 & 0 & 0 & 0 & 1 & i & 0 & 0 & 0 & 0 \\
0 & 0 & 0 & 0 & 0 & 0 & 0 & -1 & i & 0 \\
0 & 0 & 0 & 0 & 0 & 0 & 0 & 0 & 0 & \sqrt{2} \\
0 & 0 & 0 & 0 & 0 & 0 & 0 & 1 & i & 0
\end{pmatrix}
\end{array}
\end{eqnarray}

The generators used in the text should always be understood to be
related to the ones defined above as $A.T^\alpha . A^\dag$ for the
fundamental representation and $B.T^\alpha . B^\dag$ for the adjoint
representation.

\section{Gauge and Fermion Wave Functions with $h\neq0$}
\label{funcs}
\renewcommand{\theequation}{C.\arabic{equation}}
\setcounter{equation}{0}  

\renewcommand{\thefigure}{C.\arabic{figure}}
\setcounter{figure}{0}  

It should be understood in the following that all the previously
defined functions, $S_{\pm M_i}, \dot{S}_{\pm M_i},S$ and $C$ are
functions of $x_5$ and $z$, where $z=m_n$ and $m_n$ is taken to be
the mass of the particle of interest. The functions $\theta_{Gx}$
and $\theta_{Fx}$ are defined as:

\begin{eqnarray}\label{theta}
\theta_{Gx}&=&\frac{\lambda_G h f_h}{f^2_{hx}},\\
&&\nonumber\\
\theta_{Fx}&=&\frac{\lambda_F h f_h}{f^2_{hx}},
\end{eqnarray}
where,
\begin{equation}\label{fhx}
f^2_{hx}=\frac{1}{g_5^2 \int_0^{x_5} dy a^{-2}(y) }
\end{equation}

The gauge boson wave functions in the presence of the Higgs vev are
as follows:

\begin{eqnarray}
&&f_{G,\hat{a}}(x_5,h)=\left[\begin{array}{c}
\left(\cos\theta_{G_x}
C_{G,\hat{1}}+\frac{1}{\sqrt{2}}\sin\theta_{G_x}
C_{G,1_R}\right)S[x]-\frac{1}{\sqrt{2}}\sin\theta_{G_x} C_{G,1_L}C[x]\\
\\
\left(\cos\theta_{G_x}
C_{G,\hat{2}}+\frac{1}{\sqrt{2}}\sin\theta_{G_x}
C_{G,2_R}\right)S[x]-\frac{1}{\sqrt{2}}
\sin\theta_{G_x} C_{G,2_L}C[x]\\
\\
\cos\theta_{G_x}\text{
}C_{G,\hat{3}}S[x]+\frac{1}{\sqrt{2}}\sin\theta_{G_x}
\left( c_{\phi } C_{G,3_R}S[x]+\left(s_{\phi } C_{G,Y}-C_{G,3_L}\right)C[x] \right)\\
\\
S[x] C_{G,\hat{4}}\end{array}\right]\nonumber\\
&&\nonumber\\
&&\label{c.4}
\end{eqnarray}

\begin{eqnarray}
f_{G,a_L}(x_5,h)=\nonumber\\
&\left[\begin{array}{c} \frac{1}{2} \left(
\left(1+\cos\theta_{G_x}\right) C_{G,1_L}C[x]+ \left(\sqrt{2}
\sin\theta_{G_x} C_{G,\hat{1}}+\left(1-\cos\theta_{G_x}\right) C_{G,1_R}\right)S[x]\right)\\
\\
\frac{1}{2} \left( \left(1+\cos\theta_{G_x}\right)
C_{G,2_L}C[x]+\left(\sqrt{2}
\sin\theta_{G_x} C_{G,\hat{2}}+\left(1-\cos\theta_{G_x}\right) C_{G,2_R}\right)S[x] \right)\\
\\
\frac{1}{2} \left( \left(\left(1-\cos\theta_{G_x}\right) c_{\phi }
C_{G,3_R}+\sqrt{2} \sin\theta_{G_x} C_{G,\hat{3}}\right)S[x]\right.\qquad\qquad\qquad\qquad\nonumber\\
\qquad\qquad\qquad\qquad\left.+
\left(\left(1+\cos\theta_{G_x}\right)
C_{G,3_L}+\left(1-\cos\theta_{G_x}\right) s_{\phi }
C_{G,Y}\right)C[x]\right)\end{array}\right]&\nonumber\\ &&\nonumber\\
&&
\end{eqnarray}

\begin{eqnarray}
f_{G,a_R}(x_5,h)=\nonumber\\
&\left[\begin{array}{c}\frac{1}{2}
\left(\left(1-\cos\theta_{G_x}\right) C_{G,1_L}C[x] -
\left(\sqrt{2}
\sin\theta_{G_x} C_{G,\hat{1}}-\left(1+\cos\theta_{G_x}\right) C_{G,1_R}\right)S[x]\right)\\
\\
\frac{1}{2} \left(\left(1-\cos\theta_{G_x}\right) C_{G,2_L}C[x] -
\left(\sqrt{2}
\sin\theta_{G_x} C_{G,\hat{2}}-\left(1+\cos\theta_{G_x}\right) C_{G,2_R}\right)S[x]\right)\\
\\
\frac{1}{2} \left(\left(\left(1+\cos\theta_{G_x}\right) c_{\phi }
C_{G,3_R}-\sqrt{2} \sin\theta_{G_x}
C_{G,\hat{3}}\right)S[x]\right.\qquad\qquad\qquad\qquad\nonumber\\
\qquad\qquad\qquad\qquad\left.+\left(\left(1-\cos\theta_{G_x}\right)
C_{G,3_L}+\left(1+\cos\theta_{G_x}\right) s_{\phi } C_{G,Y}\right)C[x] \right)\end{array}\right]&\nonumber\\ &&\nonumber\\
&&\label{c.6}
\end{eqnarray}
\begin{eqnarray}
&&f_{G,X}(x_5,h)=\left[\begin{array}{c}c_{\phi }C_{G,Y} C[x] -s_{\phi } C_{G,3_R}S[x] \end{array}\right]\nonumber\\ &&\nonumber\\
&&\label{c.7}
\end{eqnarray}

The fermion wave functions in the presence of the Higgs vev are
given by the following:

\begin{eqnarray}\label{F1L.h}
f_{F,1,L}(x_5,h)=\nonumber\\
&\left[\begin{array}{c}C_{F,1} S_{M_1}\\
\\
\frac{1}{2} \left(\left(1+\cos\theta_{F_x}\right) C_{F,2}
S_{M_1}-\left(1-\cos\theta_{F_x}\right) C_{F,3} \dot{S}_{-M_1}-\sqrt{2} \sin\theta_{F_x} C_{F,5} S_{M_1}\right)\\
\\
\frac{1}{2} \left(\left(1+\cos\theta_{F_x}\right) C_{F,3}
\dot{S}_{-M_1}-\left(1-\cos\theta_{F_x}\right) C_{F,2} S_{M_1}-\sqrt{2} \sin\theta_{F_x} C_{F,5} S_{M_1}\right)\\
\\
C_{F,4} \dot{S}_{-M_1}\\
\\
\frac{1}{\sqrt{2}}\sin\theta_{F_x} \left(C_{F,2} S_{M_1}+ C_{F,3}
\dot{S}_{-M_1}\right)+\cos\theta_{F_x} C_{F,5} S_{M_1}
\end{array}\right]&\nonumber\\
&&\nonumber\\
&&
\end{eqnarray}

\begin{eqnarray}\label{F2R.h}
f_{F,2,R}(x_5,h)=\nonumber\\
&\left[\begin{array}{c}C_{F,6}S_{-M_2}\\
\\
\frac{1}{2} \left(\left(1+\cos\theta_{F_x}\right) C_{F,7}S_{-M_2}
-\left(1-\cos\theta_{F_x}\right) C_{F,8} S_{-M_2}-\sqrt{2} \sin\theta_{F_x} C_{F,10} \dot{S}_{M_2}\right)\\
\\
\frac{1}{2} \left(\left(1+\cos\theta_{F_x}\right) C_{F,8}S_{-M_2}
-\left(1-\cos\theta_{F_x}\right)C_{F,7} S_{-M_2} -\sqrt{2} \sin\theta_{F_x}C_{F,10} \dot{S}_{M_2} \right)\\
\\
C_{F,9}S_{-M_2}\\
\\
\frac{1}{\sqrt{2}}\sin\theta_{F_x} \left(C_{F,7}+C_{F,8}\right)
S_{-M_2}+\cos\theta_{F_x} C_{F,10} \dot{S}_{M_2}
\end{array}\right]&\nonumber\\
&&\nonumber\\
&&
\end{eqnarray}

\begin{eqnarray}\label{F3R.h}
f_{F,3,R}(x_5,h)=\nonumber\\
&\left[\begin{array}{c}\cos\theta_{F_x}C_{F,11} S_{-M_3} -\frac{i
}{\sqrt{2}}\sin\theta_{F_x}\left(C_{F,15}-C_{F,18}\right)S_{-M_3}\\
\\
\frac{1}{2} \left(\left(1+\cos\theta_{F_x}\right)
C_{F,12}-\left(1-\cos\theta_{F_x}\right) C_{F,13}-i \sin\theta_{F_x} \left(C_{F,16}-C_{F,19}\right)\right)S_{-M_3} \\
\\
\frac{1}{2} \left(\left(1+\cos\theta_{F_x}\right)
C_{F,13}-\left(1-\cos\theta_{F_x}\right) C_{F,12}-i \sin\theta_{F_x} \left(C_{F,16}-C_{F,19}\right)\right)S_{-M_3} \\
\\
\cos\theta_{F_x}\text{ }C_{F,14}S_{-M_3}-\frac{i}{\sqrt{2}}
\sin\theta_{F_x}
\left(C_{F,17}S_{-M_3} - C_{F,20}\dot{S}_{M_3}\right)\\
\\
\frac{1}{2}\text{ }\left(\left(1+\cos\theta_{F_x}\right)
C_{F,15}+\left(1-\cos\theta_{F_x}\right) C_{F,18}-i \sqrt{2} \sin\theta_{F_x} C_{F,11}\right)S_{-M_3}\\
\\
\frac{1}{2} \left(\left(1+\cos\theta_{F_x}\right)
C_{F,16}+\left(1-\cos\theta_{F_x}\right) C_{F,19}-i \sin\theta_{F_x} \left(C_{F,12}+C_{F,13}\right)\right) S_{-M_3}\\
\\
\frac{1}{2} \left(\left(1+\cos\theta_{F_x}\right)C_{F,17}S_{-M_3}+\left(1-\cos\theta_{F_x}\right)C_{F,20}\dot{S}_{M_3}-i \sqrt{2} \sin\theta_{F_x} C_{F,14} S_{-M_3}\right)\\
\\
\frac{1}{2} \left(i \sqrt{2} \sin\theta_{F_x}
C_{F,11}+\left(1-\cos\theta_{F_x}\right) C_{F,15}+\left(1+\cos\theta_{F_x}\right) C_{F,18}\right)S_{-M_3} \\
\\
\frac{1}{2} \left(i \sin\theta_{F_x}
\left(C_{F,12}+C_{F,13}\right)+\left(1-\cos\theta_{F_x}\right) C_{F,16}+\left(1+\cos\theta_{F_x}\right) C_{F,19}\right)S_{-M_3} \\
\\
\frac{1}{2} \left(i \sqrt{2}
\sin\theta_{F_x}C_{F,14}S_{-M_3}+\left(1-\cos\theta_{F_x}\right)
C_{F,17}S_{-M_3}+\left(1+\cos\theta_{F_x}\right) C_{F,20}
\dot{S}_{M_3}\right)
\end{array}\right]&\nonumber\\
&&\nonumber\\
&&
\end{eqnarray}

The boundary equations separate by charge as said previously. For
the $5/3$ charge $C_{F,1},C_{F,11},C_{F,15}$ and $C_{F,18}$ is one
coupled set given by,
\begin{eqnarray}
&2C_{F,1} \dot{S}_{M_1}[L]+M_{B_2} \left(2 \cos\left[\frac{\lambda_F h}{f_h}\right] C_{F,11}+i \sqrt{2} \sin\left[\frac{\lambda_F h}{f_h}\right] \left(C_{F,15}-C_{F,18}\right)\right) S_{-M_3}[L]=0,&\nonumber\\
&&\label{53.12}\\
&2 M_{B_2} C_{F,1} S_{M_1}[L]-\left(2 \cos\left[\frac{\lambda_F
h}{f_h}\right] C_{F,11}+i \sqrt{2}
\sin\left[\frac{\lambda_F h}{f_h}\right] \left(C_{F,15}-C_{F,18}\right)\right) \dot{S}_{-M_3}[L]=0,&\nonumber\\
&&\label{53.22}\\
& \left(i \sqrt{2} \sin\left[\frac{\lambda_F h}{f_h}\right]
C_{F,11}+\left(1+\cos\left[\frac{\lambda_F h}{f_h}\right]\right)
C_{F,15}+\left(1-\cos\left[\frac{\lambda_F h}{f_h}\right]\right) C_{F,18}\right) \dot{S}_{-M_3}[L]=0,&\nonumber\\
&&\label{53.32}\\
& \left(-i \sqrt{2} \sin\left[\frac{\lambda_F h}{f_h}\right]
C_{F,11}+\left(1-\cos\left[\frac{\lambda_F h}{f_h}\right]\right)
C_{F,15}+\left(1+\cos\left[\frac{\lambda_F h}{f_h}\right]\right) C_{F,18}\right) \dot{S}_{-M_3}[L]=0&,\nonumber\\
&&\label{53.42}
\end{eqnarray}
the equation involving $C_{F,6}$ is independent. The $-1/3$
charged fermions are also given by two sets:
$C_{F,4},C_{F,14},C_{F,17}, C_{F,2}$  which includes the bottom,
\begin{eqnarray}
&2C_{F,4} S_{-M_1}[L]+ M_{B_2} \left(2 \cos\left[\frac{\lambda_F h}{f_h}\right] C_{F,14} S_{-M_3}[L]+i \sqrt{2} \sin\left[\frac{\lambda_F h}{f_h}\right] \left(C_{F,17} S_{-M_3}[L]-C_{F,20} \dot{S}_{M_3}[L]\right)\right)=0,&\nonumber\\
&&\label{b.1}\\
&2\left(\cos\left[\frac{\lambda_F h}{f_h}\right] C_{F,14}
\dot{S}_{-M_3}[L]-M_{B_2} C_{F,4}
\dot{S}_{-M_1}[L]\right)-i\sqrt{2}
\sin\left[\frac{\lambda_F h}{f_h}\right] \left(C_{F,20} S_{M_3}[L]-C_{F,17} \dot{S}_{-M_3}[L]\right)=0,&\nonumber\\
&&\label{b.2}\\
&\left(1-\cos\left[\frac{\lambda_F h}{f_h}\right]\right) C_{F,20}
S_{M_3}[L]+\left(i \sqrt{2} \sin\left[\frac{\lambda_F h}{f_h}\right] C_{F,14}+\left(1+\cos\left[\frac{\lambda_F h}{f_h}\right]\right) C_{F,17}\right) \dot{S}_{-M_3}[L]=0,&\nonumber\\
&&\label{b.3}\\
& \left(1+\cos\left[\frac{\lambda_F h}{f_h}\right]\right) C_{F,20}
S_{M_3}[L]+\left(-i \sqrt{2}
\sin\left[\frac{\lambda_F h}{f_h}\right] C_{F,14}+\left(1-\cos\left[\frac{\lambda_F h}{f_h}\right]\right) C_{F,17}\right) \dot{S}_{-M_3}[L]=0,&\nonumber\\
&&\label{b.4}
\end{eqnarray}
and one independently $C_{F,9}$.

The set for the charge $2/3$ fermion involve $C_{F,2},C_{F,3},C_{F,5},C_{F,7},C_{F,8}$, $C_{F,10},C_{F,12},C_{F,13},C_{F,16},C_{F,19}$, which
includes the top and is given by,
\begin{eqnarray}
& M_{B_2} \left(\left(1+\cos\left[\frac{\lambda_F
h}{f_h}\right]\right) C_{F,12}-\left(1-\cos\left[\frac{\lambda_F
h}{f_h}\right]\right) C_{F,13}+i \sin\left[\frac{\lambda_F
h}{f_h}\right] \left(C_{F,16}-C_{F,19}\right)\right)
S_{-M_3}[L]&\nonumber\\
&-\left(1-\cos\left[\frac{\lambda_F h}{f_h}\right]\right) C_{F,3}
S_{-M_1}[L]+\left(\left(1+\cos\left[\frac{\lambda_F
h}{f_h}\right]\right)
C_{F,2}-\sqrt{2} \sin\left[\frac{\lambda_F h}{f_h}\right] C_{F,5}\right) \dot{S}_{M_1}[L] =0,&\nonumber\\
&&\label{t.1}\\
&M_{B_2} \left(\left(1-\cos\left[\frac{\lambda_F
h}{f_h}\right]\right) C_{F,12}-\left(1+\cos\left[\frac{\lambda_F
h}{f_h}\right]\right) C_{F,13}-i \sin\left[\frac{\lambda_F
h}{f_h}\right] \left(C_{F,16}-C_{F,19}\right)\right)
S_{-M_3}[L]&\nonumber\\
& -\left(1+\cos\left[\frac{\lambda_F h}{f_h}\right]\right) C_{F,3}
S_{-M_1}[L]+\left(\left(1-\cos\left[\frac{\lambda_F
h}{f_h}\right]\right)
C_{F,2}+\sqrt{2} \sin\left[\frac{\lambda_F h}{f_h}\right] C_{F,5}\right) \dot{S}_{M_1}[L]=0,&\nonumber\\
&&\label{t.2}
\end{eqnarray}

\begin{eqnarray}
&\sin\left[\frac{\lambda_F h}{f_h}\right] \left(C_{F,3}
S_{-M_1}[L]+M_{B_1} \left(C_{F,7}+C_{F,8}\right)
S_{-M_2}[L]+C_{F,2}
\dot{S}_{M_1}[L]\right)&\nonumber\\
&+\sqrt{2}\cos\left[\frac{\lambda_F h}{f_h}\right] \left(C_{F,5} \dot{S}_{M_1}[L]+M_{B_1} C_{F,10} \dot{S}_{M_2}[L]\right)=0,&\nonumber\\
&&\label{t.3}\\
&\sin\left[\frac{\lambda_F h}{f_h}\right] \left(-M_{B_1}
\left(C_{F,2} S_{M_1}[L]+C_{F,3}
\dot{S}_{-M_1}[L]\right)+\left(C_{F,7}+C_{F,8}\right)
\dot{S}_{-M_2}[L]\right)&\nonumber\\
&-\sqrt{2}\cos\left[\frac{\lambda_F h}{f_h}\right] \left(M_{B_1} C_{F,5} S_{M_1}[L]-C_{F,10} S_{M_2}[L]\right)=0,&\nonumber\\
&&\label{t.4}\\
&\sqrt{2} \sin\left[\frac{\lambda_F h}{f_h}\right] C_{F,10}
S_{M_2}[L]-\left(\left(1+\cos\left[\frac{\lambda_F
h}{f_h}\right]\right)
C_{F,7}-\left(1-\cos\left[\frac{\lambda_F h}{f_h}\right]\right) C_{F,8}\right) \dot{S}_{-M_2}[L]=0,&\nonumber\\
&&\label{t.5}\\
&\sqrt{2} \sin\left[\frac{\lambda_F h}{f_h}\right] C_{F,10}
S_{M_2}[L]+\left(\left(1-\cos\left[\frac{\lambda_F
h}{f_h}\right]\right)
C_{F,7}-\left(1+\cos\left[\frac{\lambda_F h}{f_h}\right]\right) C_{F,8}\right) \dot{S}_{-M_2}[L]=0,&\nonumber\\
&&\label{t.6}\\
&M_{B_2} \left(\left(\left(1+\cos\left[\frac{\lambda_F
h}{f_h}\right]\right) C_{F,2}-\sqrt{2} \sin\left[\frac{\lambda_F
h}{f_h}\right] C_{F,5}\right)
S_{M_1}[L]-\left(1-\cos\left[\frac{\lambda_F h}{f_h}\right]\right)
C_{F,3}
\dot{S}_{-M_1}[L]\right)&\nonumber\\
&-\left(\left(1+\cos\left[\frac{\lambda_F h}{f_h}\right]\right)
C_{F,12}-\left(1-\cos\left[\frac{\lambda_F
h}{f_h}\right]\right) C_{F,13}+i \sin\left[\frac{\lambda_F h}{f_h}\right] \left(C_{F,16}-C_{F,19}\right)\right) \dot{S}_{-M_3}[L]=0,&\nonumber\\
&&\label{t.7}\\
&M_{B_2} \left(\left(\left(1-\cos\left[\frac{\lambda_F
h}{f_h}\right]\right) C_{F,2}+\sqrt{2} \sin\left[\frac{\lambda_F
h}{f_h}\right] C_{F,5}\right)
S_{M_1}[L]-\left(1+\cos\left[\frac{\lambda_F h}{f_h}\right]\right)
C_{F,3}
\dot{S}_{-M_1}[L]\right)&\nonumber\\
&-\left(\left(1-\cos\left[\frac{\lambda_F h}{f_h}\right]\right)
C_{F,12}-\left(1+\cos\left[\frac{\lambda_F
h}{f_h}\right]\right) C_{F,13}-i \sin\left[\frac{\lambda_F h}{f_h}\right] \left(C_{F,16}-C_{F,19}\right)\right) \dot{S}_{-M_3}[L]=0,&\nonumber\\
&&\label{t.8}\\
&\left(i \sin\left[\frac{\lambda_F h}{f_h}\right]
\left(C_{F,12}+C_{F,13}\right)+\left(1+\cos\left[\frac{\lambda_F
h}{f_h}\right]\right)
C_{F,16}+\left(1-\cos\left[\frac{\lambda_F h}{f_h}\right]\right) C_{F,19}\right) \dot{S}_{-M_3}[L]=0,&\nonumber\\
&&\label{t.9}\\
& \left(i \sin\left[\frac{\lambda_F h}{f_h}\right] \left(C_{F,12}+C_{F,13}\right)-\left(1-\cos\left[\frac{\lambda_F h}{f_h}\right]\right) C_{F,16}-\left(1+\cos\left[\frac{\lambda_F h}{f_h}\right]\right) C_{F,19}\right) \dot{S}_{-M_3}[L]=0,&\nonumber\\
&&\label{t.10}
\end{eqnarray}

\renewcommand{\theequation}{B.\arabic{equation}}
\setcounter{equation}{0}  

\renewcommand{\thefigure}{B.\arabic{figure}}
\setcounter{figure}{0}  

\end{document}